\documentclass[10pt,a4paper]{elsarticle}
\usepackage[utf8]{inputenc}
\usepackage{amsmath}
\usepackage{amsfonts}
\usepackage{amssymb}
\usepackage{graphicx}
\usepackage{fullpage}
\usepackage{float}
\usepackage[colorlinks,allcolors=blue]{hyperref}
\usepackage{appendix}

\newcommand{\Rep}{\mathrm{Re}_{\mathrm{p}}}

\DeclareMathOperator\erf{erf}

\makeatletter
\def\ps@pprintTitle{%
  \let\@oddhead\@empty
  \let\@evenhead\@empty
  \let\@oddfoot\@empty
  \let\@evenfoot\@oddfoot
}
\makeatother

\begin{document}

\begin{frontmatter}

\title{Steady undisturbed velocity correction scheme for Euler-Lagrange simulations near planar walls}

\author[affil1]{Akshay Chandran}
\author[affil2]{Fabien Evrard}
\author[affil1]{Berend van Wachem\corref{cor1}}
\ead{berend.van.wachem@multiflow.org}

\address[affil1]{Chair of Mechanical Process Engineering, Otto-von-Guericke-Universit\"at Magdeburg, Universit\"atsplatz 2, 39106 Magdeburg, Germany}
\address[affil2]{Department of Aerospace Engineering, University of Illinois Urbana-Champaign, 104 South Wright Street, Urbana, IL 61801, United States of America}

\cortext[cor1]{Corresponding author: }

\date{}

\begin{abstract}
Euler-Lagrange (EL) point-particle simulations rely on hydrodynamic force closure models to accurately predict particle dynamics in flows. The closure 
models currently employed for dilute particle-laden flows require the undisturbed fluid velocity estimated at 
the particle center. Recovering this undisturbed velocity necessitates modeling the particle-induced disturbance on the flow. In this work, we present a new framework for velocity disturbance modeling near no-slip walls, suited for dilute gas-solid flows characterized by large Stokes numbers. For small disturbance Reynolds numbers, the velocity disturbance governing equations reduce to the steady Stokes equations. To exactly satisfy the no-slip and no-penetration boundary conditions for the velocity disturbance at the location of the planar wall, we employ the method of images to derive the Green's functions of the Stokes equations for Wendland and Gaussian force regularization kernels. An additional convolution product with a mesh-spacing dependent Gaussian kernel is performed to match the solution produced by a discrete flow solver. We verify the proposed analytical model by comparing the velocity disturbance generated on a computational mesh with the model predictions. An extension of the model for finite particle Reynolds numbers is proposed by multipying an Oseen correction factor in an unbounded domain. The correction scheme is validated using canonical test cases involving the motion of a single particle parallel and perpendicular to a wall at various Stokes numbers, particle Reynolds numbers, and particle diameter to mesh-spacing ratios. Owing to the polynomial representation of the velocity disturbance for a Wendland force regularization, the current model is computationally efficient and well-suited for large-scale, two-way coupled EL simulations.
\end{abstract}


\end{frontmatter}




\section{Introduction}

Particle-laden flows are commonplace in natural and industrial settings. From the motion of dust particles in air, 
red blood cells in capillaries, and inert silica particles used to enhance fluidization in particle beds, the list is endless. 
These flows encompass a wide range of length and time scales, which can be computationally challenging to 
resolve using computational fluid solvers. Accurate particle-resolved simulations of such real-world scenarios requires the spatial and temporal discretization of the flow 
to resolve the smallest structures around the particles and the shortest time scales governing their motion. In the dilute regime, where the particle volume 
fraction is $\mathcal{O}(10^{-3})$ or lower, these requirements become extremely demanding \cite{Balachandar2010,Mehrabadi2018}.

The Euler-Lagrange (EL) point-particle methodology provides a solution to this problem, specifically in the dilute regime \cite{Elghobashi1993,Truesdell1994,Mehrabadi2018}. EL simulations do not resolve the particle surface, 
rather represent the particles as point sources. Their motion is dictated by Newton's second law, while particle-fluid interactions are simplified using reduced-order force models that capture the particle-fluid momentum exchange. Consequently, EL simulations are ideally suited for studying the collective behaviour of a large number of particles, which remains computationally infeasible using particle-resolved direct numerical simulations \cite{Balachandar2010,Wachem2022}. The reduced force closures that model particle-fluid momentum coupling rely on accurately estimating the undisturbed fluid velocity at the particle center. The undisturbed fluid velocity corresponds to the fluid velocity that would exist in the absence of the particle. For example, in the case of a single particle sedimenting in a quiescent flow, the undisturbed velocity is exactly zero. However, conventional two-way coupled EL simulations can only access the disturbed fluid velocity. Since the particle presence induces a flow disturbance, recovering the undisturbed velocity 
requires subtracting the particle-induced disturbance on the flow. Depending on the characteristic spatial and temporal scales of the flow, various empirical and analytical disturbance schemes have been proposed over the last decades \cite{Capecelatro2013,Ireland2017,Balachandar2019,Evrard2020a,Apte2022,Horwitz2022,Balachandar2022,
Evrard2025,Pakseresht2020a,Chandran2025}.

Models that determine the velocity disturbance near walls additionally need to ensure the modeled velocity disturbance satisfies the appropriate wall boundary conditions. Various approaches have been proposed in the literature to model the velocity disturbance near planar walls. \citet{Pakseresht2020a} quantify the velocity disturbance generated by a computational fluid cell exerting the same force as a particle in the Stokes regime, incorporating empirical factors to correct for the presence of a wall, inertial effects, shape of the cell, and unsteady effects of the disturbance. \citet{Horwitz2022} use finite-volume simulations of a point forcing centered at various wall-normal distances in a channel geometry, tabulating the resulting velocity disturbance to extract the Green's functions of the steady Stokes equations in a channel geometry. Recently, \citet{Chandran2025} propose a deterministic transient velocity disturbance model that utilized the solution of the unsteady Stokes equation near planar walls \cite{Felderhof2009} to model the velocity disturbance generated by low Stokes number particles. However, the computation of the velocity disturbance required storage and tracking of fictitious fluid particles, which becomes intractable for large scale particle-laden flows. \citet{Balachandar2022} propose to satisfy no-slip conditions for the velocity disturbance by using a mirror particle across the wall. However, this model does not accurately capture the decay of the velocity disturbance towards the wall, and further only manages to enforce a slip boundary condition across the wall. \citet{Battista2019} obtain the velocity disturbance by solving an auxiliary vorticity equation without the need for any empirical factors. Although their model can be used for arbitrary particle to fluid density ratios, they also use a mirrored forcing near the boundary, making their methodology suitable for very small particle diameters compared to the local curvature of the boundary. \citet{Apte2022} propose to use an overset grid to solve a model disturbance equation in a small region around each particle. For a large number of particles, this method requires evaluation of the undisturbed velocity for every particle using an individual overset mesh, making the process computationally very expensive. In summary, most models currently available in literature either rely on empirical factors to account for presence of the wall or are too computationally expensive for practical use in large-scale particle-laden flows.

Gas-solid flows with large particle diameter to fluid length-scale ratios and with large density ratios between the two phases are charaterized by high Stokes numbers. The Stokes number is defined as the ratio of the characteristic acceleration time scale of the particle, $\tau_{\mathrm{p}}$, to that of the fluid, $\tau_{\mathrm{f}}$. In such flows, due to the significant difference in time scales, the self-induced velocity disturbance of the particle plateaus to a steady state within each $\tau_{\mathrm{p}}$. As a result, the observed velocity disturbance by the particle is steady. Currently, no analytical steady velocity disturbance correction schemes for such particles near walls exist, and it is this gap that the present work 
aims to address, in addition to developing a model with a small computational and memory footprint. We employ the method of images to derive the velocity disturbance that satisfies boundary conditions near planar walls. The method of images was originally proposed by \citet{Blake1971} for singular Stokeslets.
\citet{Ainley2008} extended the framework for a regularized 
Stokeslet for an algebraic regularization kernel, and \citet{Cortez2015} generalized it for arbitrary regularization kernels. In 
this work, we derive the system of images for a Wendland and Gaussian regularization kernels, leveraging the 
general framework developed by \citet{Cortez2015}. We apply this solution in the context of undisturbed 
velocity correction for moderate to large Stokes number particle-laden flows. The resulting expressions for the velocity disturbance are computationally cheap to evaluate, introducing only a negligible cost to an EL simulation.

The remainder of the paper is organized as follows. In Section \ref{sec:mathematical_formulation}, we derive the analytical steady velocity disturbance using the method of images. Section \ref{sec:numerical_methodology} outlines the necessary framework for the numerical implementation of the model. In Section \ref{sec:results}, we compare the obtained analytical velocity disturbance against mesh-generated velocity disturbances, and validate the present disturbance correction scheme through single particle simulations in a quiescent flow. Section \ref{sec:summary_conclusions} summarizes the study and presents our conclusions. \ref{sec:appedixa_companionblobs_gf} provides a detailed derivation of the regularization kernel pair and their corresponding Green's functions for the Wendland and Gaussian kernels used in deriving the image system. \ref{sec:appendixb_regelements_freespace} details the free-space regularized elements of the Wendland and Gaussian kernels. 
Finally, \ref{sec:appendixc_taylorseries_oseencorrection} provides a Taylor series expansion of the Oseen correction factor for a Wendland 
force regularization kernel.


\section{Mathematical formulation}
\label{sec:mathematical_formulation}

The motion of a particle in the Euler-Lagrange framework is governed by Newton's second law:

\begin{equation}
\rho_{\mathrm{p}} V_{\mathrm{p}} \frac{\mathrm{d} \mathbf{u}_{\mathrm{p}}}{\mathrm{d} t} = \mathbf{F}_{\mathrm{fluid}} + \mathbf{F}_{\mathrm{body}},
\label{eqn:newtons_secondlaw_particles}
\end{equation}

where $\rho_{\mathrm{p}}$ is the particle density, $V_{\mathrm{p}}$ is the particle volume, and $\mathbf{u}_{\mathrm{p}}$ is the center of mass velocity 
of the particle. $\mathbf{F}_{\mathrm{body}}$ represents the volumetric forces, such as gravity, acting on the particle. 
$\mathbf{F}_{\mathrm{fluid}}$ encompasses the drag, history, added mass, and undisturbed flow force contributions acting on the particle, modeled using the Maxey-Riley-Gatignol (MRG) equations \cite{Maxey1983,Gatignol1983,Bergougnoux2014}. Since the particle geometry is reduced to a single point denoting its center of mass, the description of fluid forces on the particle depends on the particle velocity and fluid flow quantities, such as velocity and pressure, interpolated to the particle center. For instance, 
the drag force on a rigid spherical particle of diameter $d_{\mathrm{p}}$, centered at $\mathbf{x}_{\mathrm{p}}$, and immersed in a fluid of density $\rho_{\mathrm{f}}$ is:

\begin{equation}
\mathbf{F}_{\mathrm{p},\mathrm{drag}} = \frac{3}{4} \frac{\rho_{\mathrm{f}} V_{\mathrm{p}} \mathrm{C}_{\mathrm{D}}}{d_{\mathrm{p}}} \lvert \lvert \tilde{\mathbf{u}}(\mathbf{x}_{\mathrm{p}}) -  \mathbf{u}_{\mathrm{p}} \rvert \rvert ( \tilde{\mathbf{u}}(\mathbf{x}_{\mathrm{p}}) -  \mathbf{u}_{\mathrm{p}}).
\label{eqn:drag_pointparticle}
\end{equation}

Here $\rho_{\mathrm{f}}$ is the fluid density, and $\mathrm{C}_{\mathrm{D}}$ represents the drag coefficient \cite{Schiller1933}. $\tilde{\mathbf{u}}(\mathbf{x}_{\mathrm{p}})$ denotes the \textit{undisturbed fluid velocity} at the particle center, defined as the fictitious fluid velocity that would exist in the absence of the particle. Due to the velocity disturbance caused by the momentum feedback from the particle on to the fluid, conventional EL simulations only have access to the \textit{disturbed fluid velocity}, $\overline{\mathbf{u}}$, at the particle location. The undisturbed velocity is then recovered by subtracting a modeled velocity disturbance, $\mathbf{u}^{\prime}$, from the disturbed fluid velocity.

Velocity disturbance models have been derived both empirically \cite{Pakseresht2020a,Horwitz2016} and theoretically in the Stokes or Oseen flow regimes \cite{Balachandar2019,Evrard2020a,Balachandar2022,Evrard2025,Chandran2025}, for unbounded and wall-bounded flows. For particles near walls, these velocity disturbance models must satisfy the no-slip and no-penetration conditions at the walls. Additionally, the choice between a steady or unsteady velocity disturbance model depends on the Stokes number of the particle-fluid system. Here, the Stokes number is defined as the ratio of the characteristic particle acceleration time scale, $\tau_{\mathrm{p}}$, to the fluid viscous time scale based on the filter kernel width, $\tau_{\nu}$. When $\tau_{\mathrm{p}} \ll \tau_{\nu}$, the timescale of momentum diffusion across the filter kernel is much larger than the particle timescale, hence one must consider the transient evolution of the particle-induced velocity disturbance in order to model it accurately \cite{Evrard2025,Pakseresht2020a,Chandran2025}. Conversely when $\tau_{\mathrm{p}} \gg \tau_{\nu}$, the timescale at which the particle-induced flow disturbance reacts to changes in the magnitude and direction of forcing is much smaller than the particle timescale, hence a steady disturbance model may be used. In the following subsection, we introduce a steady velocity 
disturbance model for particles near planar walls, based on the method of images \cite{Blake1971,Ainley2008,Cortez2015}.

\subsection{Steady velocity disturbance model}

In this work, we aim to address a gap in the literature regarding EL simulations of gas-solid flows 
characterized by large Stokes numbers. In such scenarios, it can be safely assumed the particle-induced velocity disturbance 
develops on a much shorter timescale than the particle timescale, making it suitable to use a quasi-steady description for the  
velocity disturbance. Additionally, in the limit of the $\mathrm{Re}^{\prime} = \lVert \mathbf{u}^{\prime} \rVert \delta / \nu_{\mathrm{f}} \ll 1$, 
the velocity disturbance equations takes the form of the Stokes equations. Here, $\mathrm{Re}^{\prime}$ is the Reynolds number based 
on the disturbance velocity scale, $\mathbf{u}^{\prime}$ and filter kernel length, $\delta$ in a fluid of kinematic viscosity, $\nu_{\mathrm{f}}$. The governing equations for 
$\mathbf{u}^{\prime}$ can then be expressed as follows:

\begin{equation}
-\nabla p^{\prime} + \mu_{\mathrm{f}} \nabla^2 \mathbf{u}^{\prime} + \mathbf{F} g(\mathbf{x}) = 0, \quad \nabla \cdot \mathbf{u}^{\prime} = 0,
\label{eqn:steadyStokes_velocitydisturbanceeqns}
\end{equation}

where $p^{\prime}$ denotes the pressure, $\mu_{\mathrm{f}}$ represents the dynamic viscosity of the fluid, and $\mathbf{F}$ 
is the force exerted by the particle on the fluid, obtained from reduced force models. The function $g(\mathbf{x})$ represents the 
regularization kernel of the forcing $\mathbf{F}$, with a kernel support radius of $\delta$. For particles near 
planar walls, $\mathbf{u}^{\prime}$ must additionally satisfy the no-slip and no-penetration boundary 
conditions at the wall.

For a singular forcing near a planar wall, i.e., when $g(\mathbf{x})$ is the Dirac delta function, 
\citet{Blake1971} presents the solution of the Stokes equations near a planar wall using a system of images, 
analytically achieving an exact cancellation of the velocity at the wall. Blake's system of images comprises of a Stokeslet located at the particle center, its image positioned on the opposite side of the wall, and an auxiliary equation for the remainder velocity, solved in Fourier space. In particle-laden flows of practical relevance, however, the momentum 
feedback from the particle is spatially distributed. This can be accomplished using a regularization kernel of finite 
support such as a Wendland or a truncated Gaussian kernel. However, extending the method of images for 
regularization kernels with finite support does not inherently result in an exact cancellation of the velocity at the wall. 
It has been shown that for regularization kernels with finite support, there exists a companion kernel for every 
$g(\mathbf{x})$ such that the velocity boundary conditions are satisfied \cite{Cortez2015}. In the present 
section, we utilize the general framework developed by \citet{Cortez2015} to derive the Stokeslet image system for Wendland and Gaussian regularized forcings, providing an analytical solution for 
$\mathbf{u}^{\prime}$ that exactly satisfies the velocity boundary conditions at the wall.

\begin{figure}[H]
\centering
\includegraphics[scale=1.0]{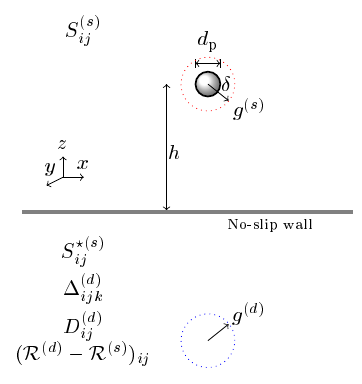}
\caption{Schematic showing the image system of the regularized elements of the free-space Stokes equations and their positions relative to the wall, 
for a particle of diameter $d_{\mathrm{p}}$ located at a distance $h$ away from the planar wall, with a force regularization kernel radius of $\delta$. Here, $S_{ij}^{(s)}$ and $S_{ij}^{\star,(s)}$ represent the free-space regularized Stokeslet kernel centered at a distance $h$ above and below the no-slip wall respectively, and are derived using the regularization kernel $g^{(s)}$. $\Delta_{ijk}^{(d)}$, $D_{ij}^{(d)}$, and $\mathcal{R}_{ij}^{(d)}$ represent free-space regularized elements derived using the secondary blob $g^{(d)}$, and are centered at a distance $h$ below the no-slip wall.}
\label{fig:imagesystem_moi}
\end{figure}

Figure \ref{fig:imagesystem_moi} illustrates a schematic of the Stokeslet image system and their locations with respect 
to the no-slip wall for a particle of diameter $d_{\mathrm{p}}$ located at a distance $h$ away from a planar wall, 
with a regularization kernel radius of $\delta$. The dotted red circle represents the regularization kernel 
$g(\mathbf{x})$, henceforth referred to as the Stokeslet kernel, with a superscript $(s)$ denoting regularized 
elements derived from this kernel. The dotted blue circle corresponds to the regularization kernel of the companion 
kernel as prescribed in \citet{Cortez2015}, and detailed in \ref{subsec:appedixa_companionblobs_gf_wendland} 
and \ref{subsec:appedixa_companionblobs_gf_gaussian}. This companion kernel is henceforth referred to as the dipole 
kernel, and a superscript $(d)$ is used to denote regularized elements derived using this kernel. In the schematic, $S_{ij}^{(s)}$ denotes the free-space Stokeslet kernel and is the only regularized element located above the wall. $S_{ij}^{\star(s)}$ 
represents the mirror image of $S_{ij}^{(s)}$ across the wall, $\Delta_{ijk}^{(d)}$ is the regularized doublet, 
$D_{ij}^{(d)}$ is the regularized dipole, and $\mathcal{R}_{ij}^{(d)}$ denotes the regularized rotlet. 
$S_{ij}^{\star,(s)}$, $\Delta_{ijk}^{(d)}$, $D_{ij}^{(d)}$, and $\mathcal{R}_{ij}^{(d)}$ are all centered at a distance 
of $2h$ from $S_{ij}$ across the wall.

\begin{figure}[H]
\centering
\includegraphics[scale=1.0]{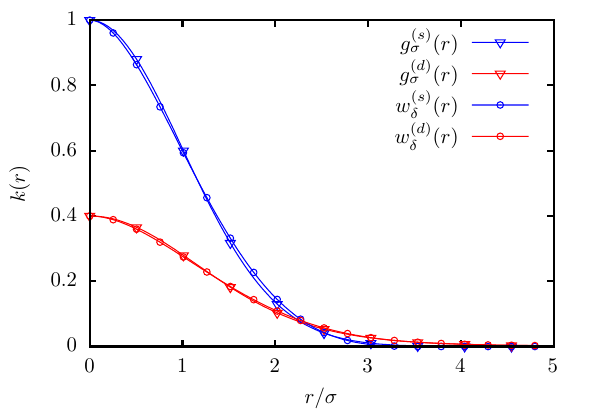}
\caption{Stokeslet blobs, $g_{\sigma}^{(s)}$ and $w_{\delta}^{(s)}$, and the corresponding companion dipole blobs, $g_{\sigma}^{(d)}$ and $w_{\delta}^{(d)}$ for Gaussian and Wendland regularization kernels. Here, $\delta = \sigma \sqrt{9 \pi/2}$.}
\label{fig:gaussian_stokesletdipoleblobs}
\end{figure}

Equations \eqref{eqn:wendlandkernel_stokesletblob_dipoleblob} and \eqref{eqn:gaussiankernel_stokesletblob_dipoleblob} 
show the Stokeslet and dipole kernels for a Wendland kernel with a compact support radius $\delta$ and a Gaussian kernel 
with a standard deviation $\sigma$ respectively. For further details of this derivation, the reader is referred to \ref{subsec:appedixa_companionblobs_gf_wendland} and \ref{subsec:appedixa_companionblobs_gf_gaussian}. 
Figure \ref{fig:gaussian_stokesletdipoleblobs} compares the Stokeslet and dipole kernels 
presented in Equations \eqref{eqn:wendlandkernel_stokesletblob_dipoleblob} and 
\eqref{eqn:gaussiankernel_stokesletblob_dipoleblob}. To enable a meaningful comparison between the Wendland and Gaussian kernels, the 
radius of the Wendland support is related to the standard deviation of the Gaussian by $\delta = \sigma \sqrt{9 \pi/2}$ \cite{Evrard2020a}.

\begin{align}
\begin{split}
w_{\delta}^{(s)} (r) &= \frac{21}{2 \pi \delta^3} \biggl\{ \begin{array}{ll} (4r/\delta + 1)(1 - r/\delta)^4  & r \leq \delta, \\ 0 & r > \delta. \end{array},
\\
w_{\delta}^{(d)} (r) &= \frac{21}{5 \pi \delta^3}  \biggl\{ \begin{array}{ll} 1 - \frac{50}{7} (r/\delta)^2 + \frac{25}{2} (r/\delta)^3 - \frac{25}{3} (r/\delta)^4 + 2 (r/\delta)^5  & r \leq \delta, \\ \frac{1}{42} (\delta/r)^5 & r > \delta. \end{array}.
\end{split}
\label{eqn:wendlandkernel_stokesletblob_dipoleblob}
\end{align}

\begin{align}
\begin{split}
g_{\sigma}^{(s)} (r) &= \frac{1}{(2 \pi \sigma^2)^{3/2}} \exp \left( \frac{-r^2}{2 \sigma^2} \right),
\\
g_{\sigma}^{(d)} &= \frac{1}{\sqrt{2} \pi^{3/2} r^5 \sigma} \Biggl[ -r (r^2 + 3 \sigma^2) \exp \left( \frac{-r^2}{2 \sigma^2} \right) + 3 \sigma^3 \sqrt{\frac{\pi}{2}} \erf \left( \frac{r}{\sqrt{2} \sigma} \right) \Biggr].
\end{split}
\label{eqn:gaussiankernel_stokesletblob_dipoleblob}
\end{align}

\begin{figure}[H]
\includegraphics[scale=1.5]{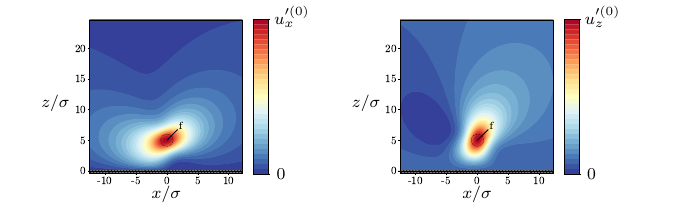}
\caption{The disturbance velocity fields, $u_{x}^{\prime}$ and $u_{z}^{\prime}$ generated due to the Gaussian image system for a force, $\mathbf{f} = \{3 \pi \mu_{\mathrm{f}} d_{\mathrm{p}},0,3 \pi \mu_{\mathrm{f}} d_{\mathrm{p}}\}$ 
at a distance of $h = 5 \sigma$ away from the wall located at $z = 0$. Here, the colourbars are normalized with the corresponding maximum velocities, $u_{x}^{\prime(0)}$ and $u_{z}^{\prime(0)}$ respectively in the domain at the center of forcing. The dashed black circle represents has a radius equal to the standard deviation of the Gaussian regularization kernel, and the gray rectangle denotes a no-slip, no-penetration wall.}
\label{fig:gaussiandipole_uxvelocity}
\end{figure}

The regularized elements required for the image system are straightforward to derive and are detailed in \ref{sec:appendixb_regelements_freespace}. The resulting expression for the Stokeslet image system is given as follows 
\cite{Cortez2015}:

\begin{equation}
S_{ij}^{\mathrm{IM}} = (S_{ij}^{\star(s)} - S_{ij}^{(s)}) - \Biggl[ 2 h \rho_{mj} \left( \Delta_{i3m}^{(d)} + \frac{h}{2} D_{im}^{(d)} \right) + 2 h (\mathcal{R}^{(d)} - \mathcal{R}^{(s)})_{im} \epsilon_{jm3} \Biggr],
\label{eqn:imagesystem_stokeslet_main}
\end{equation}

where $\rho_{mj} = (\delta_{mj} - 2 \delta_{m3} \delta_{j3})$. Here, $\delta_{ab}$ is the Kronecker delta function.
$S_{ij}^{\mathrm{IM}}$ represents the superposed Stokeslet kernel that enforces zero velocity boundary conditions 
at the wall for a Stokeslet kernel $g(\mathbf{x})$ centered at an arbitrary distance $h$ away from the wall. 
Since each of the regularized elements is a solution of Equation \eqref{eqn:steadyStokes_velocitydisturbanceeqns}, 
the superposed kernel $S_{ij}^{\mathrm{IM}}$ is also a valid solution owing to the linearity of the Stokes equations. 
Additionally, since we are interested in the physical half-space above the wall, i.e., $z > 0$, the image system 
only includes the regularization kernel of the forcing in this half-space, while providing a solution that 
satisfies the required velocity boundary conditions at $z = 0$. The particle-induced velocity disturbance on the 
fluid is then simply: 

\begin{equation}
u_i^{\prime} = S_{ij}^{\mathrm{IM}} f_j, 
\label{eqn:veldisturbance_kernelforcing_contraction}
\end{equation}

where $f_j$ is the force exerted by the particle on the fluid. It can be shown that for any arbitrary point on the 
wall, $S_{ij}^{\mathrm{IM}} (z = 0) = 0$, thereby enforcing $u_i^{\prime} = 0$. Figure 
\ref{fig:gaussiandipole_uxvelocity} shows the $u_x^{\prime}$ and $u_{z}^{\prime}$ velocity response for a forcing directed along
 $\mathbf{f} = \{3 \pi \mu_{\mathrm{f}} d_{\mathrm{p}},0,3 \pi \mu_{\mathrm{f}} d_{\mathrm{p}}\}$ at a distance of $h = 5 \sigma$ away from the wall located at $z = 0$. The maximum velocity distubance, $u_{x}^{\prime(0)}$ and $u_{z}^{\prime(0)}$ at the center of the forcing are used to normalize the field.

\section{Numerical methodology}
\label{sec:numerical_methodology}

The velocity disturbance model presented in Equation \eqref{eqn:veldisturbance_kernelforcing_contraction} and 
detailed in \ref{sec:appendixb_regelements_freespace} is easy to implement, 
and adds a fractional computational cost compared to other velocity disturbance correction models near walls 
\cite{Chandran2025}. For the Wendland kernel, the 
resulting expressions for the velocity disturbance takes the form of a polynomial of order 7, making the 
evaluation of $S_{ij}^{\mathrm{IM}}$ computationally efficient. In this section, we detail the modifications required for the model to accurately evaluate the velocity disturbance generated on a discrete mesh. Additionally, we verify the 
modified analytical velocity disturbance against mesh-generated velocity disturbances. We also highlight the 
importance of using the mesh-generated velocity disturbance for the case of a particle settling towards its terminal 
velocity.

\subsection{Volume-averaged Stokeslet image system}
\label{subsec:volume_avg_imagesystem}

In a finite-volume fluid solver, the disturbed fluid velocity obtained on the computational mesh represents the mesh-averaged 
velocity disturbance at the cell centers of the mesh. Consequently, directly using an analytical model for the velocity 
disturbance tends to overpredict the mesh-generated velocity disturbance, unless the mesh is very fine. Recovering the mesh-averaged disturbance 
requires averaging the analytical velocity disturbance with a filter that is a function of the mesh-spacing, 
$\Delta x$. Mathematically, this can be achieved via a convolution of the analytical Eulerian fields with a tophat 
kernel, $\mathcal{K}_q$, where $q$ is the filter length scale dictated solely by $\Delta x$. The governing 
equations for the mesh-averaged velocity disturbance can then be expressed as follows:

\begin{equation}
\mathcal{K}_q \star \left( -\nabla p^{\prime} + \mu_{\mathrm{f}} \nabla^2 \mathbf{u}^{\prime} + \mathbf{F} g(\mathbf{x})  \right) = 0.
\label{eqn:meshavg_stokeseqns_convoperation}
\end{equation}

Here, $\star$ represents the real-space convolution integral in the three-dimensional space occupied by the fluid 
above the wall. The linearity of the Stokes equations permits us to write Equation \eqref{eqn:meshavg_stokeseqns_convoperation} as,

\begin{equation}
-\nabla p^{\prime}_{q} + \mu_{\mathrm{f}} \nabla^2 \mathbf{u}_{q}^{\prime} + \mathbf{F} g_{q}(\mathbf{x}) = 0.
\label{eqn:meshavg_stokeseqns}
\end{equation}

Here, $p^{\prime}_q$ and $\mathbf{u}_{q}^{\prime}$ are the mesh-averaged pressure and velocity fields respectively. 
$g_q(\mathbf{x})$ is the kernel obtained on convolution of the force regularization kernel $g(\mathbf{x})$ 
(such as the Gaussian or Wendland kernel) with the mesh-filter kernel $\mathcal{K}_q$. Obtaining $g_q(\mathbf{x})$, and subsequently 
$p^{\prime}_q$ and $\mathbf{u}_{q}^{\prime}$ is non-trivial 
when $\mathcal{K}_q$ is a tophat kernel. In this study, we assume $\mathcal{K}_q$ to be either a Gaussian or a 
Wendland kernel with a mesh dependent length-scale, enabling analytical evaluation of Equation \eqref{eqn:meshavg_stokeseqns}. 
For this purpose, the kernel $\mathcal{K}_q$ is chosen to be a Gaussian in the present work. The convoluted kernel 
$g_q(\mathbf{x})$ is another Gaussian kernel with a modified regularization radius.
The convolution of a Gaussian force regularization kernel of standard deviation $\sigma$ and another Gaussian with 
standard deviation of $q$, gives us a resulting Gaussian kernel with a 
standard deviation $\gamma = (\sigma^2 + q^2)^{1/2}$. By comparing the mesh-generated velocity disturbance 
against the analytical velocity disturbance obtained with a kernel length-scale of $\gamma$, we determine the 
relationship of $q$ with $\Delta x$, the mesh-spacing. This relationship is expressed as:

\begin{equation}
q = k_0 \left( \frac{3}{4 \pi} \right)^{1/3} \Delta x.
\label{eqn:meshfilter_lengthscale}
\end{equation}

Comparison of the mesh-averaged analytical solution with corresponding finite-volume simulations of a fixed forcing 
located at various wall-normal distances, we determined that with a proportionality factor of $k_0 = 1.8$, the analytical velocity field reproduces 
the mesh-generated velocity disturbance with minimal error. At first glance, representing a tophat kernel with a Gaussian 
or a Wendland might seem like a simplistic choice. However, we show in the subsequent sections that the proportionality 
factor of $k_0 = 1.8$ captures the mesh-averaged velocity disturbance with good accuracy. It is also important to note that 
approximating the mesh averaging using a Gaussian kernel provides a better estimate of the velocity disturbance on the mesh 
than would otherwise be obtained from the analytical velocity disturbance from Equation \eqref{eqn:veldisturbance_kernelforcing_contraction}. 
Since EL simulations with regularized forcing require the force regularization to be well-resolved, directly applying the 
analytical velocity disturbance from Equation \eqref{eqn:veldisturbance_kernelforcing_contraction} would still provide a 
good approximation of the mesh-generated velocity disturbance. EL simulations where the particle momentum source is transferred 
to a single mesh cell, i.e., PSIC, requires a mesh-averaging since $\delta / \Delta x < 1$.

\begin{figure}[H]
\centering
\includegraphics[scale=0.79]{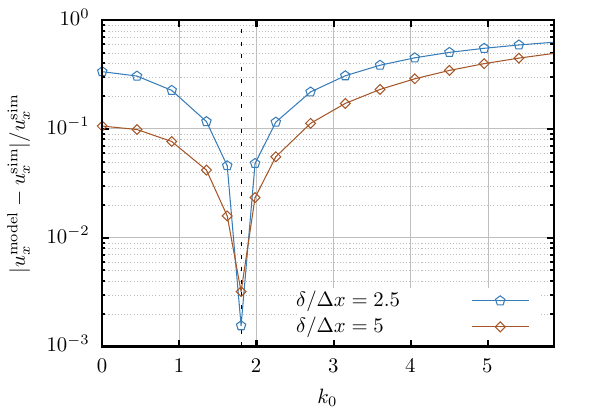}
\caption{Relative error between the model velocity, $u_x^{\mathrm{model}}$ compared to the discrete velocity from a finite-volume solver, $u_x^{\mathrm{sim}}$ measured at the center of the regularized forcing for two different resolutions of the regularization radius of $\delta/ \Delta x = \{2.5,5\}$. The black dashed line represents the location of $k_0 = 1.8$.}
\label{fig:relativeerror_k0}
\end{figure}

Figure \ref{fig:relativeerror_k0} shows the relative error in the $x$-directional velocity measured at the center of the regularized forcing from the proposed model, $u_x^{\mathrm{model}}$, and from a finite-volume simulation, $u_x^{\mathrm{sim}}$. Two different resolutions of the regularization radius $\delta/ \Delta x = \{2.5,5\}$ are shown. The minimum error is obtained around $k_0 = 1.8$, indicated by the black dashed line.

Obtaining the velocity disturbance, $\mathbf{u}^{\prime}$, as the Gaussian convoluted with the analytical velocity disturbance model from Equations \eqref{eqn:imagesystem_stokeslet_main} and \eqref{eqn:veldisturbance_kernelforcing_contraction}, the undisturbed velocity, $\tilde{\mathbf{u}}$, can now be defined as:

\begin{equation}
\tilde{\mathbf{u}} = \overline{\mathbf{u}} - \mathbf{u}^{\prime}.
\label{eqn:undisturbed_velocity_defn}
\end{equation}

Here, $\overline{\mathbf{u}}$ is the disturbed velocity obtained at the location of the particle interpolated from the computational fluid mesh in an Euler-Lagrange simulation.

\subsection{Comparison with discrete solution}

\begin{figure}[H]
\centering
\includegraphics[scale=0.79]{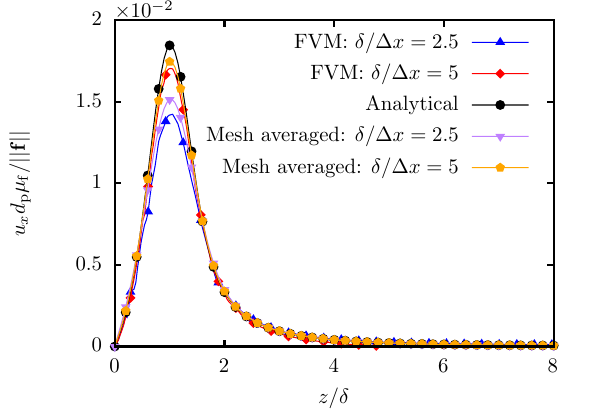}
\includegraphics[scale=0.79]{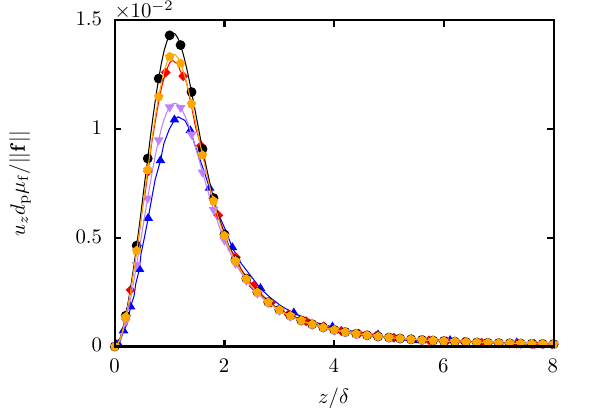}
\caption{Normalized $u_{x}$ and $u_{z}$ velocity field generated due to a Wendland regularized forcing at a distance of $h = \delta$ away from the wall using the Finite-volume solver, \texttt{MultiFlow} \cite{Denner2020} for two different mesh resolutions of $\delta/\Delta x = \{2.5, 5\}$. The obtained discrete solutions are compared against the expected analytical solution from the image system of a Wendland regularization forcing obtained from Equation \eqref{eqn:imagesystem_stokeslet_main}.}
\label{fig:wendland_uxuzvelocity_discretemodelcomparison_hdelta}
\end{figure}

\begin{figure}[H]
\centering
\includegraphics[scale=0.79]{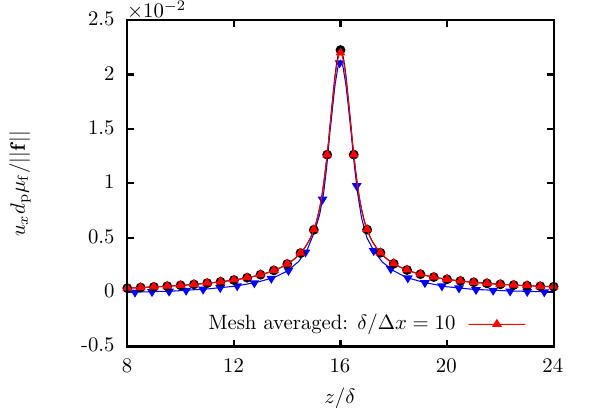}
\includegraphics[scale=0.79]{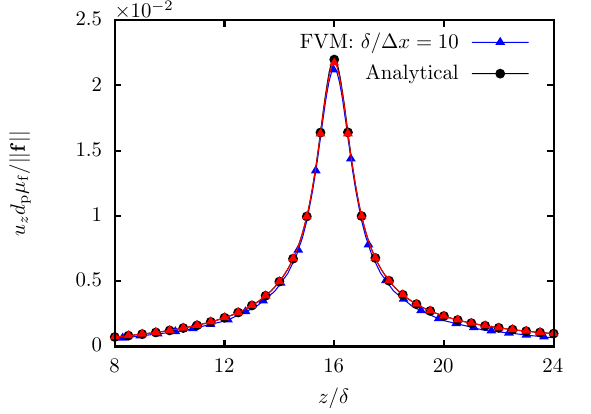}
\caption{Normalized $u_{x}$ and $u_{z}$ velocity field generated due to a Wendland regularized forcing at a distance of $h = 16 \delta$ away from the wall using the Finite-volume solver, \texttt{MultiFlow} \cite{Denner2020} for a mesh resolutions of $\delta/\Delta x = 10$. The obtained discrete solutions are compared against the expected analytical solution from the image system of a Wendland regularization forcing obtained from Equation \eqref{eqn:imagesystem_stokeslet_main}.}
\label{fig:wendland_uxuzvelocity_discretemodelcomparison_h16delta}
\end{figure}

In this subsection, we compare the discrete solution obtained from a finite-volume (FVM) solver against its analytical 
counterpart from Equation \eqref{eqn:imagesystem_stokeslet_main}. The FVM simulations are performed by varying 
the resolution of the force regularization kernel with respect to the mesh-spacing spacing $\Delta x$. We perform 
simulations for two different resolutions of $\delta/\Delta x = \{2.5,5 \}$ for a particle positioned very close to the wall, 
i.e., $h/\delta = 1$, and with a resolution of $\delta/\Delta x = 10$ for a particle far away from the wall, i.e., $h/\delta = 16$. A constant forcing $\mathbf{f} = \{3 \pi \mu_{\mathrm{f}} d_{\mathrm{p}},0,3 \pi \mu_{\mathrm{f}} d_{\mathrm{p}}\}$ 
is applied, and the steady $u_x$ (wall-parallel) and $u_z$ (wall-normal) velocity components are measured alone the wall-normal 
direction $z$. Here, $\mu_{\mathrm{f}}$ is the dynamic viscosity of the fluid, and $d_{\mathrm{p}}$ is the diameter 
of the particle. A Wendland regularization kernel is employed in all cases. The simulations are conducted using the 
second-order space and time accurate finite-volume fluid solver, \texttt{MultiFlow} \cite{Denner2020}. As shown in Figure \ref{fig:wendland_uxuzvelocity_discretemodelcomparison_hdelta}, the discrete solution of 
the Stokes equations converges to the analytical solution from Equation \eqref{eqn:imagesystem_stokeslet_main}, convoluted with a Gaussian approximation of the mesh filter using $k_0 = 1.8$ for the Wendland kernel, for both the $u_x$ and $u_z$ components of velocity. Here, the label FVM signifies simulations performed using a finite-volume solver, and the label ``Mesh averaged'' indicates the analytical velocity disturbance volume-averaged on a mesh of size $\Delta x$ using the method prescribed in the Section \ref{subsec:volume_avg_imagesystem}. A similar convergence can be observed 
in Figure \ref{fig:wendland_uxuzvelocity_discretemodelcomparison_h16delta} for a particle located at $h/\delta = 16$. In the present section and the rest, we normalize the wall-normal distance, $h$ by the length scale of the force regularization kernel, $\delta$. For a diffusion dominated velocity disturbance equation as in Equation \eqref{eqn:meshavg_stokeseqns}, $\delta$ is the predominant length scale.

\begin{figure}[H]
\centering
\includegraphics[scale=1]{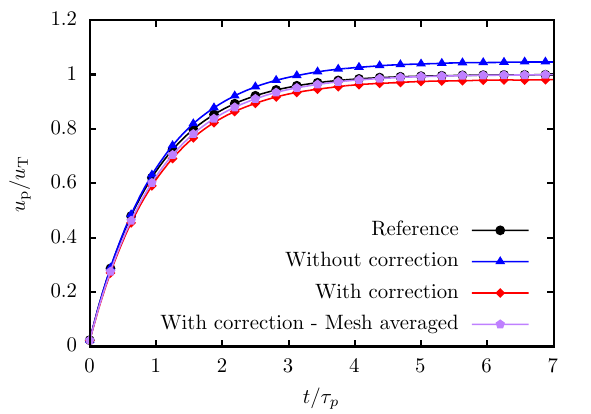}
\caption{Comparison of normalized settling velocity of a particle moving parallel to a wall using the steady correction, steady correction filtered with an approximation for the mesh filter, without any correction, and a reference one-way coupled simulation. All simulations are performed at $\Rep = 0.01$, $d_{\mathrm{p}}/\Delta x = 0.1$, $\mathrm{St} = 1.6$, and $h/\delta = 1$.}
\label{fig:ux_settling_Re0p01_hdelta_meshfilter}
\end{figure}

To evaluate the effectiveness of the correction for moving particles, we simulate a particle settling under constant external acceleration, e.g., in a constant gravitational field in a quiescent fluid at a distance of $h = \delta$ away from the wall. We perform simulations with reference one-way coupled, without any correction, with correction - without any mesh-averaging, and with correction - with a Gaussian approximation for the mesh filter. All cases are conducted at $\Rep = 0.01$, $d_{\mathrm{p}}/\Delta x = 0.1$, and Stokes number, $\mathrm{St} = 1.6$. In Figure \ref{fig:ux_settling_Re0p01_hdelta_meshfilter}, they are compared, and it can be observed that using a correction scheme without any mesh-averaging results in an overestimation of the particle-induced disturbance on the flow, leading to an underestimation of the particle settling velocity, as indicated by the curve with red diamond markers. In contrast, simulations with the $k_0 = 1.8$ approximation of the Gaussian filter are in good agreement with the velocity disturbance at the particle center, indicated by the purple hexagonal markers.

In the next section we validate the present steady correction scheme from Equation \eqref{eqn:veldisturbance_kernelforcing_contraction} used with the approximate mesh averaging from Section \ref{subsec:volume_avg_imagesystem} on the motion of particles parallel and perpendicular to a planar no-slip wall, governed by Equation \eqref{eqn:newtons_secondlaw_particles}.


\section{Results}
\label{sec:results}

In this section we validate the current model with moving particles near a wall across a range of particle Reynolds numbers, Stokes numbers, and particle diameter to mesh-spacing ratios. For finite $\Rep$, we use an additional Oseen correction factor derived for a Wendland regularization kernel in an unbounded domain \cite{Evrard2025}, extending the applicability of the present model for finite Reynolds numbers.  We rigorously delineate the limits of the present scheme, highlighting the nondimensional parameter space where the present correction performs optimally, and regimes where new correction schemes are required for dilute particle-laden flows near walls. 

\subsection{Particle settling parallel to a wall in a quiescent flow}
\label{subsec:results_parallelsettling}

In the section, we validate the current velocity disturbance model with the motion of a single rigid spherical particle moving parallel to a 
planar wall in a quiescent flow, under the influence of a gravitational force. The particle acceleration is dictated by the 
imbalance between the gravitational, drag, and buoyancy forces acting on it. At steady state, these forces are in balance, 
resulting in the particle attaining a terminal velocity. For particles in a dilute particle-laden channel flow, it is common to observe 
particles accelerating parallel to the wall under the influence of convective and viscous fluid forces. Accurately reproducing the correct particle 
trajectory can therefore be crucial in predicting reliable particle statistics in such flows \cite{Horwitz2022}. In two-way 
coupled EL simulations, the momentum imparted by the particle on the fluid results in the acceleration of the fluid in its 
vicinity along the direction of the particle motion. This leads to a lower slip velocity of the particle, and consequently a 
smaller drag force. In the absence of any velocity disturbance correction models, this underprediction of the particle drag results in the overestimation of the particle settling velocity \cite{Balachandar2019,Evrard2020a}. In scenarios where the particle imparted 
momentum on the fluid cells in its vicinity is larger, for instance when the particle diameter to mesh-spacing ratio, $d_{\mathrm{p}}/\Delta x = 1$ 
the overestimation of the particle settling velocity can have an error up to 40\% \cite{Evrard2021}. This is due to the increasing 
inertia of the particles relative to the fluid cells in its vicinity.

The nondimensional parameter characterizing the ratio of the particle to fluid time scales is the Stokes number, 
$\mathrm{St} = \tau_{\mathrm{p}} / \tau_{\mathrm{f}}$. Here, $\tau_{\mathrm{p}} = \rho_{\mathrm{p}} d_{\mathrm{p}}^2 / 18 \mu_{\mathrm{f}}$ is 
defined as the time scale of particle motion, and $\tau_{\mathrm{f}} = \rho_{\mathrm{f}} \delta^2 / \mu_{\mathrm{f}}$ is defined as 
the viscous time scale of the fluid. For large $\mathrm{St}$, i.e., $\tau_{\mathrm{f}} \ll \tau_{\mathrm{p}}$, the 
particle experiences a steady flow in its neighbourhood. In the context of velocity disturbance correction, a steady 
velocity disturbance model is thus appropriate for capturing the mesh-generated velocity disturbance in large Stokes number 
particle-laden flows. Another nondimensional parameter of interest is the particle Reynolds number, defined as:

\begin{equation}
\Rep = \frac{\rho_{\mathrm{f}} \lvert \mathbf{u}^{(\mathrm{fs})} \rvert d_{\mathrm{p}}}{\mu_{\mathrm{f}}},
\label{eqn:particleReynoldsnumber_defn}
\end{equation}

where

\begin{equation}
\mathbf{u}^{(\mathrm{fs})} = \frac{(\rho_{\mathrm{p}} - \rho_{\mathrm{f}})}{\rho_\mathrm{p}} \tau_{\mathrm{p}} \mathbf{g}
\label{eqn:unboundedsettlingvelocity_defn}
\end{equation}

is the settling velocity of the particle under an external acceleration $\mathbf{g}$ in an unbounded domain.

To validate the present model, we perform simulations with varying Stokes numbers, particle Reynolds numbers, and 
particle diameter to mesh-spacing ratios, $d_{\mathrm{p}}/\Delta x$. All simulations are performed at 
a filter radius to mesh-spacing ratio, $\delta/\Delta x = 2.5$ on a uniformly sized fluid mesh. Drag correlations for a 
particle moving parallel to a planar wall as a function of the particle separation from the wall, $h$, are provided by 
\citet{Zeng2009} and are used in the present study. The choice of the force correlation, however, does not affect the accuracy 
of the predicted velocity disturbance by the model. For all cases presented in this subsection, one-way coupled simulations 
are used as reference and for normalization, as it represents the ideal scenario with a quiescent, undisturbed fluid velocity. 
A trilinear interpolation scheme is employed to obtain the disturbed velocity and the model velocity disturbance at the particle 
center.

\begin{figure}[H]
\centering
\includegraphics[scale=1]{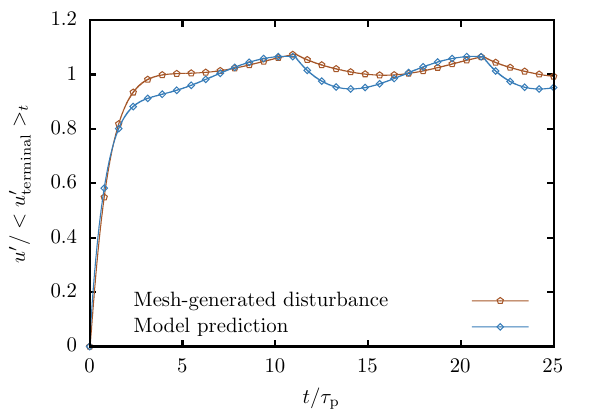}
\caption{Comparison of normalized velocity disturbance generated on a computational mesh against the predicted velocity 
disturbance from the steady disturbance model for a particle settling parallel to a wall at $\Rep = 0.01$, 
$d_{\mathrm{p}}/\Delta x = 1$, $\mathrm{St} = 1.6$, and $h/\delta = 1$. Here $<u^{\prime}_{\mathrm{terminal}}>_{t}$ is the temporally averaged velocity disturbance from the finite-volume solver for $t/\tau_{\mathrm{p}} > 8$.}
\label{fig:meshgenerated_model_velocitydisturbance_comparison}
\end{figure}

In Figure \ref{fig:meshgenerated_model_velocitydisturbance_comparison}, the velocity disturbance generated on the 
computational fluid mesh using a finite-volume solver is shown in comparison with the velocity disturbance predicted by the 
analytical steady correction model. Here, $<u^{\prime}_{\mathrm{terminal}}>_{t}$ denotes the temporally averaged velocity 
disturbance from the finite-volume solver for $t/\tau_{\mathrm{p}} > 8$. It is observed that, even with a crude approximation for the 
volume-averaging, as detailed in Section \ref{subsec:volume_avg_imagesystem}, the model captures the mesh-generated velocity disturbance. The error between the model predictions and the mesh-generated velocity disturbance tends to decrease 
for lower $d_{\mathrm{p}}/\Delta x$ cases. The oscillations observed in the velocity disturbance are due to the use of a trilinear interpolation scheme to obtain the fluid velocity at the particle center.

\begin{figure}[H]
\centering
\includegraphics[scale=1]{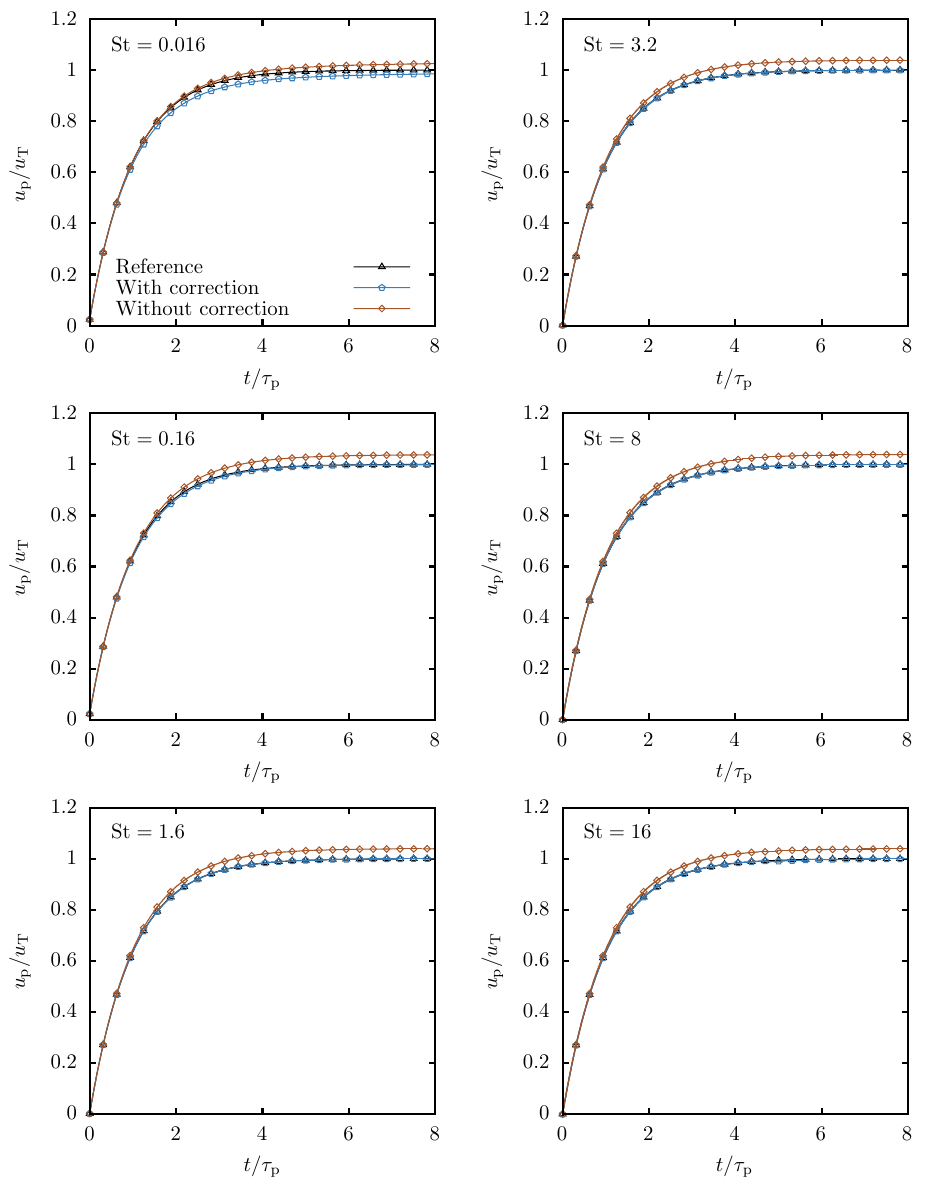}
\caption{Normalized settling velocity of a particle at various Stokes numbers at $\Rep = 0.01$, $d_{\mathrm{p}}/\Delta x = 0.1$, and $h/\delta = 1$. Comparisons are performed with the present steady correction, without any correction, and a reference one-way coupled simulation.}
\label{fig:parallelsettling_dpbydx0p1}
\end{figure}

\begin{figure}[H]
\centering
\includegraphics[scale=1]{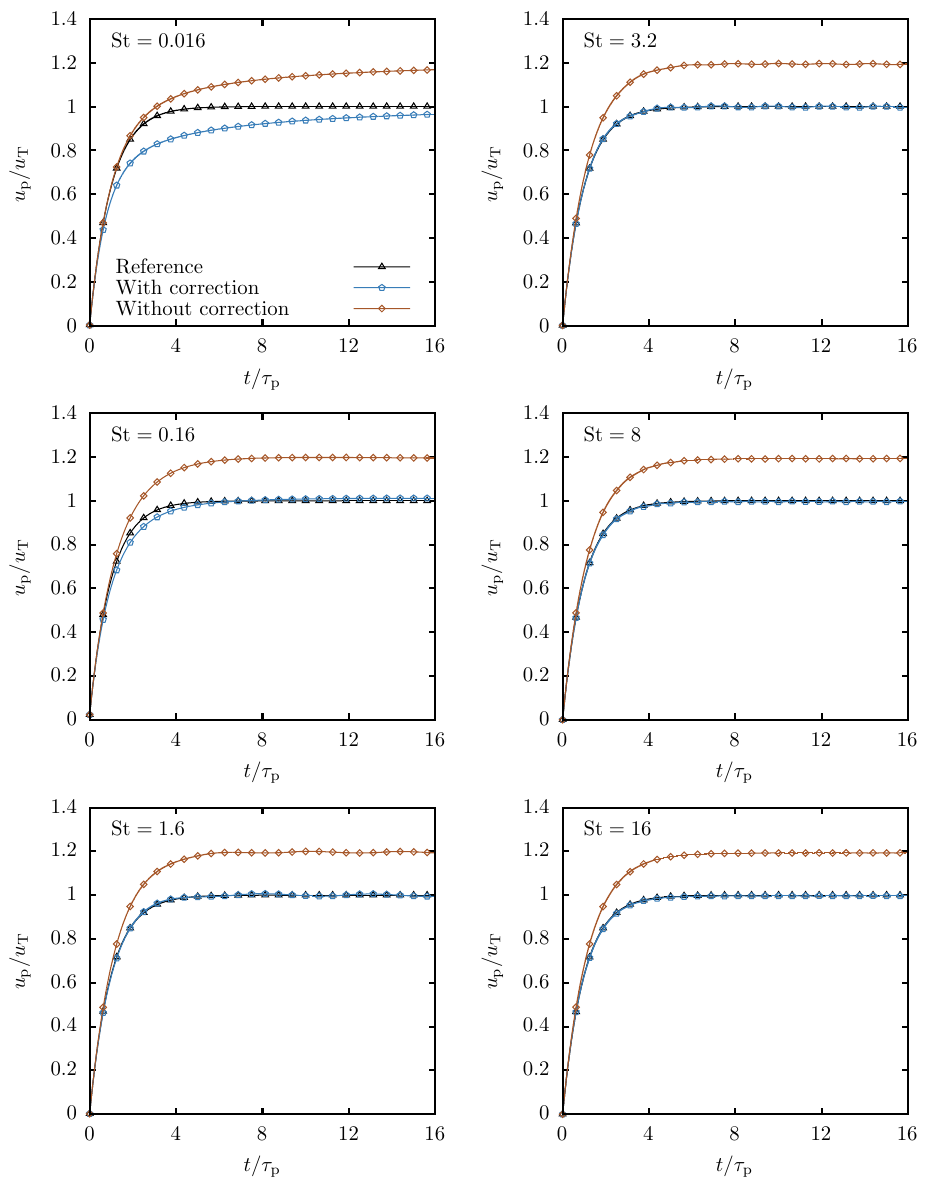}
\caption{Normalized settling velocity of a particle at various Stokes numbers at $\Rep = 0.01$, $d_{\mathrm{p}}/\Delta x = 0.5$, and $h/\delta = 1$. Comparisons are performed with the present steady correction, without any correction, and a reference one-way coupled simulation.}
\label{fig:parallelsettling_dpbydx0p5}
\end{figure}

\begin{figure}[H]
\centering
\includegraphics[scale=1]{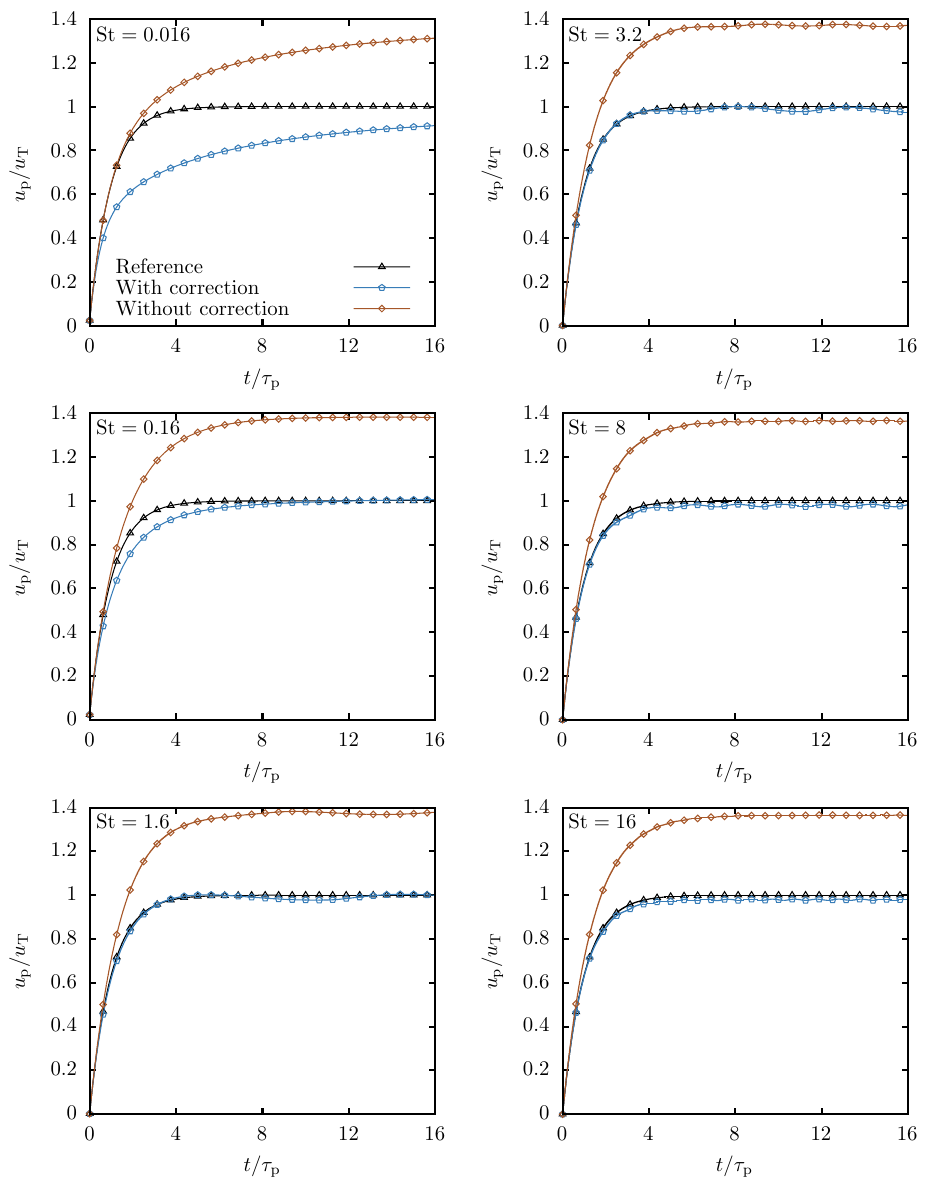}
\caption{Normalized settling velocity of a particle at various Stokes numbers at $\Rep = 0.01$, $d_{\mathrm{p}}/\Delta x = 1$, and $h/\delta = 1$. Comparisons are performed with the present steady correction, without any correction, and a reference one-way coupled simulation.}
\label{fig:parallelsettling_dpbydx1}
\end{figure}

Figures \ref{fig:parallelsettling_dpbydx0p1}, \ref{fig:parallelsettling_dpbydx0p5}, and \ref{fig:parallelsettling_dpbydx1} show 
the settling velocity of a particle moving parallel to a planar wall at a distance of $h/\delta = 1$ with a numerical resolution of 
$d_{\mathrm{p}}/\Delta x = 0.1, 0.5$, and $1$, respectively. Each case is performed at $\Rep = 0.01$ for 
six different Stokes numbers, $\mathrm{St} = \{0.016, 0.16, 1.6, 3.2, 8, 16 \}$. Due to the large momentum diffusion length scale 
and the higher drag forces acting on the particle at low $\Rep$, velocity disturbance correction schemes are particularly important 
in this flow regime. It can be observed that simulations with the present correction at $\mathrm{St} = 0.016$ and $0.16$ for all cases show a significant deviation from 
the reference one-way coupled simulations before the particle attains a terminal velocity. This deviation is a consequence of the transient 
nature of the velocity disturbance at the particle center at low Stokes numbers. Mitigating these errors requires an unsteady 
velocity disturbance model \cite{Pakseresht2020a,Chandran2025}, which are, for such cases, significantly more accurate but also computationally expensive. For the $d_{\mathrm{p}}/\Delta x = 0.1$ case, an average error of 
4\% is observed in the terminal velocity for the simulations without any correction. Comparatively, the simulations with an 
undisturbed velocity correction exhibit an error of 0.02\%. Similarly, for $d_{\mathrm{p}}/\Delta x = 0.5$, simulations from above 
$\mathrm{St} = 1.6$ without any correction predict an error of 20\% in the terminal velocity, compared to 0.7\% with the correction. For simulations with a resolution of $d_{\mathrm{p}}/\Delta x = 1$, with particles above $\mathrm{St} = 1.6$ exhibit an error 37\% and 1.5\%, without and with correction, respectively.

It is observed that the present model is versatile enough to capture the steady disturbance generated by particles for arbitrary 
$d_{\mathrm{p}}/\Delta x$. Since the generated velocity disturbance by the particle is steady, the effect of the particle diameter 
on the velocity disturbance is manifested via the particle force on the fluid, $\mathbf{f}$, and not on the Green's function kernel, 
$S_{ij}^{\mathrm{IM}}$. The velocity disturbance generated by particles of various diameters can be captured by adjusting the 
$d_{\mathrm{p}}/\Delta x$ dependent forcing, $\mathbf{f}$. Based on the simulations presented above, the steady velocity disturbance model 
is adept in capturing the particle-induced disturbance on the flow for $\mathrm{St} \geq 0.16$. Particle-laden flows with lower 
Stokes numbers require an unsteady velocity disturbance model \cite{Chandran2025}.

\subsection{Effect of particle Reynolds number}
\label{subsec:re_dependence}

The velocity disturbance model derived previously is only valid in the limit of low Reynolds numbers. 
Owing to the nonlinear nature of the disturbance governing equations at larger Reynolds numbers, the 
analytical derivation of the velocity disturbance is not possible. Alternatively, approximating the 
disturbance governing equations with the Oseen equations provides an analytic correction factor to 
the existing steady disturbance model presented in Equation \eqref{eqn:imagesystem_stokeslet_main}. 
The Oseen approximation linearizes the nonlinear convection term in the disturbance governing equations by 
treating the convective velocity to be a constant. Although it is difficult to provide this 
correction factor for the complete spatial velocity field, this factor has been previously shown to be 
derived for the velocity at the location of the particle in an unbounded domain. \citet{Balachandar2019} derived this 
analytic correction factor for a Gaussian kernel. \citet{Evrard2020a} later extended this for a 
Wendland regularization kernel. In the present work, we use the Oseen correction factor presented in 
\citet{Evrard2025}, and is shown here:

\begin{equation}
	\Psi_{\text{Os}} (\text{Re}_{\delta}) = 7 \left( \frac{1}{\text{Re}_{\delta}} - \frac{6}{\text{Re}_{\delta}^2} + \frac{30}{\text{Re}_{\delta}^3} - \frac{120}{\text{Re}_{\delta}^4} + \frac{360}{\text{Re}_{\delta}^5} - \frac{720}{\text{Re}_{\delta}^6} + \frac{720}{\text{Re}_{\delta}^7}\left( 1 - \exp\left(-\text{Re}_{\delta}\right) \right)\right).
	\label{eqn:oseen_correctionfactor}
\end{equation}

Here, $\mathrm{Re}_{\delta} = \rho_{\mathrm{f}} \lVert \tilde{\mathbf{u}} \rVert \delta / \mu_{\mathrm{f}}$ 
is the Reynolds number based on the undisturbed velocity scale and the filter length scale. 
In \ref{sec:appendixc_taylorseries_oseencorrection}, we present a Taylor series expansion of 
$\Psi_{\text{Os}} (\text{Re}_{\delta})$, and is used in the present work. This expansion is applicable 
for $\text{Re}_{\delta} \leq 10$. The velocity disturbance generated by the particle is now redefined as:

\begin{equation}
u_i^{\prime} = S_{ij}^{\mathrm{IM}} \Psi_{\text{Os}} (\text{Re}_{\delta}) f_j.
\label{eqn:oseen_corrected_veldisturbance}
\end{equation}

In the small Reynolds number limit, $\Psi_{\text{Os}} \to 1$, and the Stokes limit velocity disturbance in Equation \eqref{eqn:veldisturbance_kernelforcing_contraction} is recovered.

\begin{figure}[H]
\centering
\includegraphics[scale=1]{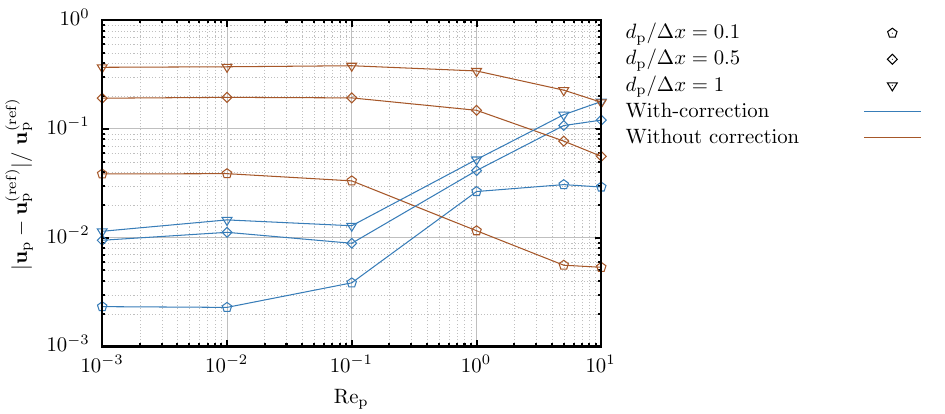}
\caption{Relative error of the particle settling velocity against a reference one-way coupled simulation at various $\Rep$, for $d_{\mathrm{p}}/\Delta x = \{0.1,0.5,1\}$, and $\mathrm{St} = 1.6$. $\delta / \Delta x = 2.5$ and $h/\delta = 1$ was maintained for all simulations. 
Here ,$\mathbf{u}_{\mathrm{p}}$ is the particle settling velocity for with or without correction, and 
$\mathbf{u}_{\mathrm{p}}^{(\mathrm{ref)}}$ represents the particle settling velocity from a one-way 
coupled simulation.}
\label{fig:re_dependence_relativeerror_dpbydx}
\end{figure}

We test the modified velocity disturbance scheme on the case of a particle settling parallel to the wall 
in a quiescent flow at $\Rep = \{10^{-3},10^{-2},10^{-1},1,5,10\}$. The simulations are performed for particle 
diameter to mesh-spacing ratios of $d_{\mathrm{p}} / \Delta x = \{0.1,0.5,1\}$ with a Stokes number of 
$\mathrm{St} = 1.6$. All simulations are performed on a regularization radius to mesh-spacing ratio of 
$\delta / \Delta x = 2.5$ and the distance of the particle center from the wall of $h/\delta = 1$ was maintained. The obtained error in the particle settling velocity are compared against 
reference one-way coupled simulations, both for simulations with and without the present correction. 
In Figure \ref{fig:re_dependence_relativeerror_dpbydx} we observe that using the steady correction with 
the Oseen factor results in upto 4\% error until $\Rep = 1$, after which simulations without any correction 
are seen to perform better in predicting the particle settling velocity. This is due to the overestimation 
of the velocity disturbance at the particle center with the steady Oseen model. Simultaneously, the 
magnitude of the particle-induced velocity disturbance on the fluid is significantly lower at larger 
$\Rep$ partially due to a decreasing drag coefficient. At $\Rep > 10$, there is only an insignificant 
effect of the self-disturbance of the particle, making any undisturbed velocity correction schemes 
mostly redundant. Based on the observed trends from Figure \ref{fig:re_dependence_relativeerror_dpbydx}, 
the linear steady Oseen correction model presented in this work should be employed upto a $\Rep = 1$, 
after which a nonlinear velocity disturbance model becomes necessary. There exists an intermediate 
range $1 < \Rep < 10$ for which this nonlinear model would ideally be suited.

\subsection{Particle falling perpendicular to a wall in a quiescent flow}

As a final validation case of the present model, we investigate the motion of a single rigid spherical particle 
moving normal to the wall under an external acceleration. The particle is initialized at a wall-normal 
separation, $l = h/d_{\mathrm{p}} - 0.5 = 7$, and is subjected to a gravitational force in the wall-normal direction, eventually bringing the particle to rest when $h/d_{\mathrm{p}} = 0.5$. To 
capture the lubrication forces acting on the particle, the drag force experienced by the particle is a 
function of the wall-normal separation, $l$. The drag coefficient expressing this relationship is as follows:

\begin{equation}
\mathrm{C_{\mathrm{D}}}^{\perp} = \frac{24}{\mathrm{Re_p}} f(l),
\end{equation}

where

\begin{equation}
f(l) = \left\{ \begin{array}{ll} 1 + \left( \frac{0.562}{1 + 2l} \right) & l > 1.38 ~~\textnormal{\cite{Brenner1961}} , \\ \\ \frac{1}{2l} \left(1 + 0.4 l \log(\frac{1}{2l}) + 1.94 l \right) & l < 1.38 ~~\textnormal{\cite{Cox1967}}. \end{array}\right .
\end{equation}

Here $f(l)$ corrects the unbounded drag coefficient for the separation of the particle from the 
planar wall. 

\begin{figure}[H]
\centering
\includegraphics[scale=1]{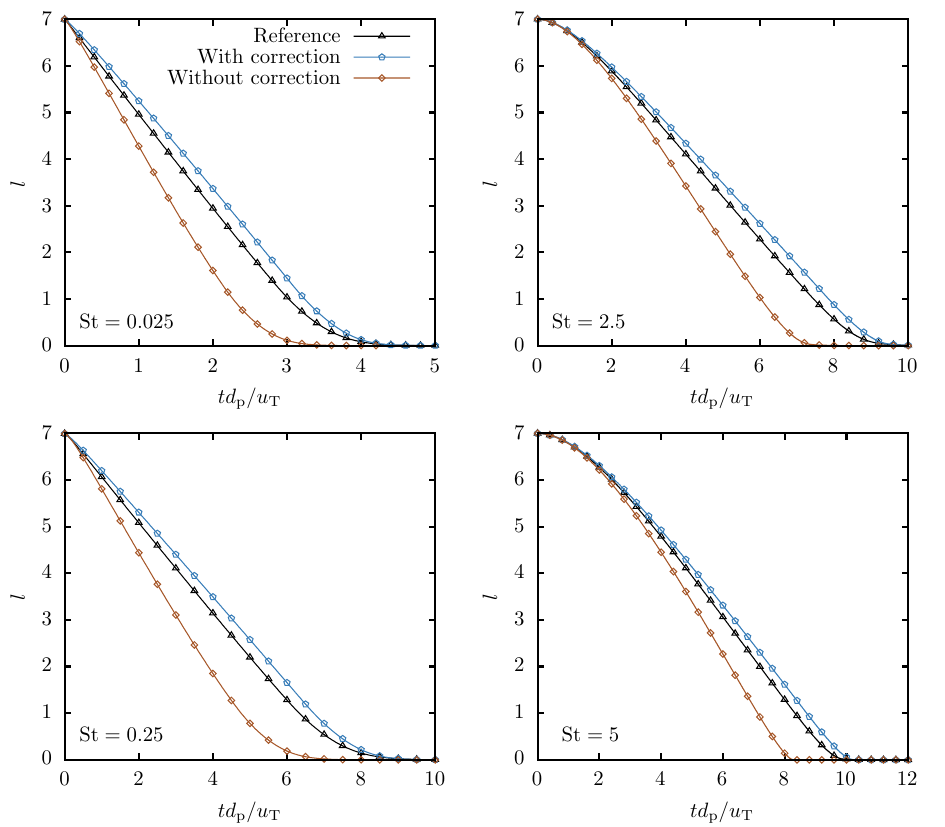}
\caption{Normalized separation, $l = h/d_{\mathrm{p}} - 0.5$ of a particle at various Stokes numbers at $Re_{\mathrm{p}} = 0.1$, $d_{\mathrm{p}}/\Delta x = 1$, and $\delta/\Delta x = 2$. Comparisons are performed with the present steady correction, without any correction, and a reference one-way coupled simulation.}
\label{fig:perpendicularsettling_normalizedseparation}
\end{figure}

For the present case, we discretize the regularization kernel radius with $\delta / \Delta x = 2$.
Due to the specification of a boundary condition along the wall-normal direction, sharp spatial velocity 
gradients are not well-resolved on a coarser mesh, requiring higher resolution of the force regularization kernel. Solving the Stokes equations on such a discrete mesh 
leads to unresolved computation of pressure and viscous terms, resulting in an inaccurate calculation of 
the flow field.

\begin{figure}[H]
\centering
\includegraphics[scale=1]{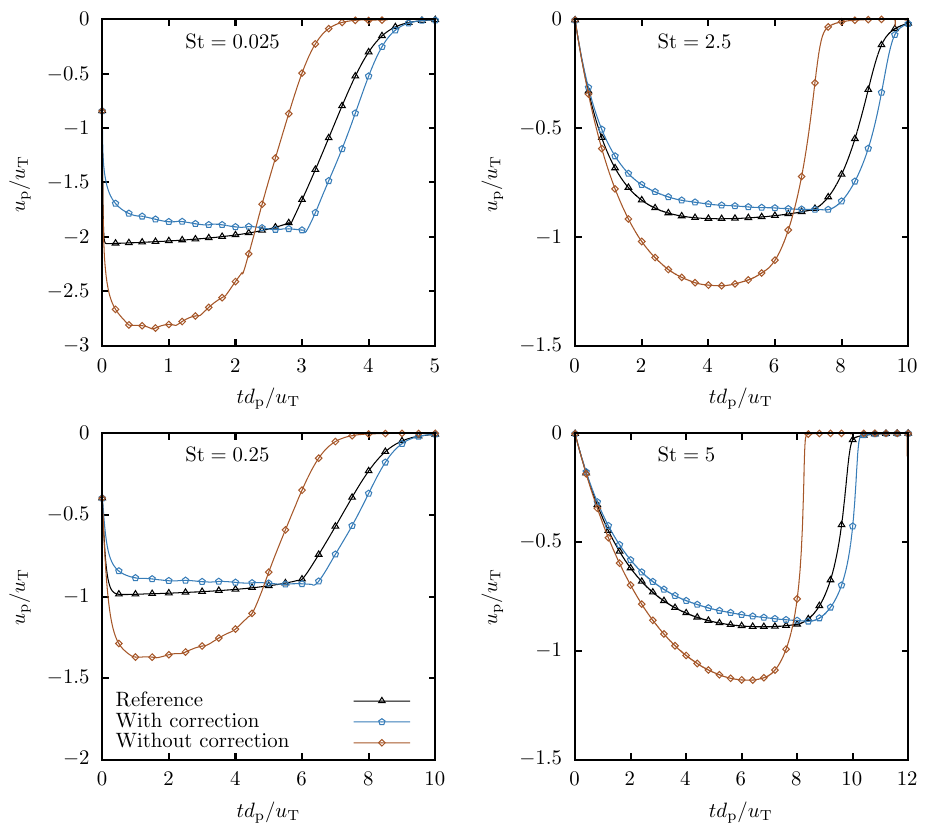}
\caption{Normalized velocity of a particle at various Stokes numbers at $Re_{\mathrm{p}} = 0.1$, $d_{\mathrm{p}}/\Delta x = 1$, and $\delta/\Delta x = 2$. Comparisons are performed with the present steady correction, without any correction, and a reference one-way coupled simulation.}
\label{fig:perpendicularsettling_normalizedvelocity}
\end{figure}

We maintain $\Rep = 0.1$ for all the simulations. Here, $\Rep$ is defined based on the velocity scale in 
the unbounded domain from Equation \eqref{eqn:unboundedsettlingvelocity_defn}. The Stokes numbers simulated 
in this case range from $\mathrm{St} = \{0.025,0.25,2.5,5\}$ for a particle diameter to mesh-spacing ratio of 
$d_{\mathrm{p}}/\Delta x = 1$. All simulations are compared against corresponding two-way coupled 
simulations without any undisturbed velocity correction, and one-way coupled simulations treated as 
reference. By comparing the normalized separation, $l$  of the particle in Figure 
\ref{fig:perpendicularsettling_normalizedseparation}, it can be seen that simulations using the present 
correction scheme predict the particle trajectory better than cases that do not use any correction. 
A similar trend can be observed for the normalized particle velocity, $u_{\mathrm{p}}/u_{\mathrm{T}}$ in 
Figure \ref{fig:perpendicularsettling_normalizedvelocity}. Additionally, similar to Section 
\ref{subsec:results_parallelsettling}, the error of the corrected simulations decrease with increasing 
Stokes numbers.

\section{Summary and conclusions}
\label{sec:summary_conclusions}

Force closure models used in Euler-Lagrange point particle simulations depend on the accurate estimation of the undisturbed fluid velocity at the particle center for predicting the motion of particles in a fluid flow. In this work, we propose a novel undisturbed velocity correction scheme for large Stokes number particle-laden flows near planar walls. The method relies on the method of images to quantify the velocity disturbance generated by a regularized forcing near a planar wall, satisfying the no-slip and no-penetration wall boundary conditions. We derive a system of regularized elements achieving the exact cancellation of velocity at the wall for Gaussian and Wendland regularization kernels, utilizing the general framework proposed by \citet{Cortez2015}. For the Wendland kernel, the system of images is represented as a seventh order polynomial, making the evaluation of the velocity disturbance computationally inexpensive, and suitable for large-scale particle-laden flows. Additionally, the method does not require storage of any predefined coefficients or fictitious particles, and therefore has a very limited memory cost associated with evaluating the velocity disturbance. We further propose an approximation of the filtering imposed by the mesh discretization in a Finite-volume solver. This is achieved by approximating the tophat kernel of the mesh filter using a Gaussian kernel with a mesh-spacing-dependent standard deviation $q$. The resulting velocity disturbance can be analytically evaluated by modifying the standard deviation of the force regularization kernel $\sigma$ as $\gamma = (\sigma^2 + q^2)^{1/2}$. Finally, we extend the model for moderate particle Reynolds numbers by incorporating a multiplicative Oseen correction factor derived for free-space.

We validate the present methodology with the motion of a rigid spherical particle moving parallel and perpendicular to the wall at various particle Reynolds numbers ($\Rep$), Stokes numbers ($\mathrm{St}$), and particle diameter to mesh-spacing ratios ($d_{\mathrm{p}}/\Delta x$). Our results show that simulations incorporating the proposed correction yield an estimated error at least one order of magnitude smaller than those predicted by simulations without any correction. We also extend the velocity disturbance model to the finite $\Rep$ regime using a multiplicative Oseen correction factor derived in an unbounded domain. We further propose that the present scheme is ideally suited for simulations where $\Rep < 1$, beyond which it tends to overestimate the velocity disturbance due to the assumption of Stokes flow in the derivation of the velocity disturbance model. Additionally, based on simulations performed at various $\mathrm{St}$, it is identified that the present scheme is most effective for $\mathrm{St} > 1$, where transient effects of the velocity disturbance are negligible. Finally, the current scheme is also versatile to handle arbitrary particle diameter to mesh-spacing ratios. We further test the model on the motion of a particle normal to a wall, where we observe similar trends, and a better prediction of the particle trajectory and velocity compared to simulations without any correction.

The fundamental advantage of the present model is the inexpensive evaluation of the self-disturbance of the particle owing to the polynomial representation of the analytical Green's function kernel for a Wendland forcing regularization. As a natural extension and possible real-life application, the present model can be readily implemented for particle-laden channel flows where the particle-induced self-disturbance is governed only by a single wall.
\section*{Acknowledgements}
This project has received funding from the Deutsche Forschungsgemeinschaft (DFG, German Research Foundation), 
grant numbers 457515061 and  457509672.
\appendix
\section{Companion blobs and Greens functions of the Wendland and Gaussian kernels}
\label{sec:appedixa_companionblobs_gf}


In this section we derive the primary and secondary blobs, referred to as the Stokeslet and dipole blobs respectively, for a Wendland and Gaussian regularization kernel.

\subsection{Wendland kernel}
\label{subsec:appedixa_companionblobs_gf_wendland}

We define the Stokeslet blob as the Wendland kernel with radius $\delta$ as follows:

\begin{equation}
w_{\delta}^{(s)} (r) = \frac{21}{2 \pi \delta^3} \left\{ \begin{array}{ll} (4r/\delta + 1)(1 - r/\delta)^4  & r \leq \delta, \\ 0 & r > \delta. \end{array}\right.
\label{eqn:wendlandkernel_stokesletblob}
\end{equation}

The Green's function for this Stokeslet blob is obtained as the solution of $\Delta G_{s} = w_{\delta}^{(s)}$. For a radially symmetric blob, we obtain:

\begin{equation}
G_{s} = \frac{1}{4 \pi \delta} \left\{ \begin{array}{ll}  3 (r/\delta)^7 - 15 (r/\delta)^6 + 28 (r/\delta)^5 - 21 (r/\delta)^4 + 7 (r/\delta)^2  & r \leq \delta, \\ 3 - (\delta/r) & r > \delta. \end{array}\right.
\label{eqn:greensfunction_stokesletblob_wendland}
\end{equation}

Similarly, the biharmonic function is obtained as the solution of $\Delta B_{s} = G_{s}$ is:

\begin{equation}
B_{s} = \frac{\delta}{120 \pi} \left\{ \begin{array}{ll}  (r/\delta)^9 - \frac{25}{4} (r/\delta)^8 + 15 (r/\delta)^7 - 15 (r/\delta)^6 + \frac{21}{2} (r/\delta)^4  & r \leq \delta, \\ 15 (r/\delta)^2 - (\delta/r) - 15 (r/\delta) + (25/4) & r > \delta. \end{array}\right.
\label{eqn:biharmonicfunction_stokesletblob_wendland}
\end{equation}

The dipole blob, $w_{\delta}^{(d)}$ is obtained as \cite{Cortez2015}:

\begin{equation}
w_{\delta}^{(d)} (r) = \frac{21}{5 \pi \delta^3} \left\{ \begin{array}{ll} 1 - \frac{50}{7} (r/\delta)^2 + \frac{25}{2} (r/\delta)^3 - \frac{25}{3} (r/\delta)^4 + 2 (r/\delta)^5  & r \leq \delta, \\ \frac{1}{42} (\delta/r)^5 & r > \delta. \end{array}\right.
\label{eqn:dipoleblob_wendland}
\end{equation}

The corresponding Green's function for the dipole blob is obtained as the solution of $\Delta G_{d} = w_{\delta}^{(d)}$, which for a radially symmetric blob is:

\begin{equation}
G_{d} = \frac{1}{20 \pi \delta} \left\{ \begin{array}{ll}  3 (r/\delta)^7 - \frac{50}{3} (r/\delta)^6 + 35 (r/\delta)^5 - 30 (r/\delta)^4 + 14 (r/\delta)^2  & r \leq \delta, \\ -5 (\delta/r) + \frac{1}{3} (\delta/r)^3 + 10 & r > \delta. \end{array}\right.
\label{eqn:greensfunction_dipoleblob_wendland}
\end{equation}

\subsection{Gaussian kernel}
\label{subsec:appedixa_companionblobs_gf_gaussian}

We define the Stokeslet blob with the Gaussian kernel with standard deviation $\sigma$ as follows:

\begin{equation}
g_{\sigma}^{(s)} (r) = \frac{1}{(2 \pi \sigma^2)^{3/2}} \exp \left( \frac{-r^2}{2 \sigma^2} \right).
\label{eqn:gaussiankernel}
\end{equation}

The Green's function for this dipole blob is obtained as the solution of $\Delta G_{s} = g_{\sigma}^{(s)}$. For a radially symmetric blob, we obtain:

\begin{equation}
G_{s} = -\frac{1}{4 \pi r} \erf \left( \frac{r}{\sqrt{2} \sigma} \right).
\label{eqn:greensfunction_stokesletblob_gaussian}
\end{equation}

Similarly, the biharmonic function is obtained as the solution of $\Delta B_{d} = G_{d}$ is:

\begin{equation}
B_{s} = - \frac{1}{4 \pi} \Biggl[ \frac{\sigma}{\sqrt{2} \pi} \exp \left( \frac{-r^2}{2 \sigma^2} \right)  + \frac{(r^2 + \sigma^2)}{2 r} \erf \left( \frac{r}{\sqrt{2} \sigma} \right) \Biggr].
\label{eqn:biharmonicfunction_stokesletblob_gaussian}
\end{equation}

The dipole blob, $g_{\sigma}^{(d)}$ is obtained as \cite{Cortez2015}:

\begin{equation}
g_{\sigma}^{(d)} = \frac{1}{\sqrt{2} \pi^{3/2} r^5 \sigma} \Biggl[ -r (r^2 + 3 \sigma^2) \exp \left( \frac{-r^2}{2 \sigma^2} \right) + 3 \sigma^3 \sqrt{\frac{\pi}{2}} \erf \left( \frac{r}{\sqrt{2} \sigma} \right) \Biggr].
\label{eqn:dipoleblob_gaussian}
\end{equation}

The corresponding Green's function for the Stokeslet blob is obtained as the solution of $\Delta G_{d} = g_{\sigma}^{(d)}$, which is for a radially symmetric blob:

\begin{equation}
G_{d} = - \frac{1}{4 \pi^{3/2}} \Biggl[ \frac{\sqrt{2} \sigma}{r^2} \exp \left( \frac{-r^2}{2 \sigma^2} \right)  + \frac{\sqrt{\pi} (r^2 - \sigma^2)}{r^3} \erf \left( \frac{r}{\sqrt{2} \sigma} \right) \Biggr].
\label{eqn:greensfunction_dipoleblob_gaussian}
\end{equation}


\section{Regularized elements in free space}
\label{sec:appendixb_regelements_freespace}

The image system for a regularized Stokeslet resulting in zero velocity at the wall is $u_i^{\prime} = S_{ij}^{\mathrm{IM}} f_j$ \cite{Cortez2015}, where:

\begin{equation}
S_{ij}^{\mathrm{IM}} = (S_{ij}^{\star(s)} - S_{ij}^{(s)}) - \Biggl[ 2 h \rho_{mj} \left( \Delta_{i3m}^{(d)} + \frac{h}{2} D_{im}^{(d)} \right) + 2 h (\mathcal{R}^{(d)} - \mathcal{R}^{(s)})_{im} \epsilon_{jm3} \Biggr],
\label{eqn:imagesystem_stokeslet}
\end{equation}

where $\rho_{mj} = (\delta_{mj} - 2 \delta_{m3} \delta_{j3})$. Here $\mathbf{S}$ and $\mathbf{S}^{\star}$ are the regularized Stokeslets, $\mathbf{D}$ is a regularized dipole, $\mathbf{\Delta}$ is a regularized doublet, and $\boldsymbol{\mathcal{R}}$ is a regularized rotlet.

\subsection{Regularized Stokeslet}

Defining the Stokeslet velocity field as $u_i = S_{ij} f_j = [H_1 (r) \delta_{ij} + H_2 (r) x_i x_j] f_j$. Here,

\begin{equation}
H_1 (r) = \frac{1}{r} \frac{\mathrm{d}B_s (r)}{\mathrm{d}r} - G_s (r) \quad \textnormal{and} \quad H_2(r) = \frac{1}{r^3} \left( r \frac{\mathrm{d}^2B_s (r)}{\mathrm{d}r^2} - \frac{\mathrm{d}B_s (r)}{\mathrm{d}r} \right).
\label{eqn:regstokeslet_h1h2_defn}
\end{equation}

For a Wendland kernel, substituting $G_s$ and $B_s$ from Equation \eqref{eqn:greensfunction_stokesletblob_wendland} and \eqref{eqn:biharmonicfunction_stokesletblob_wendland}, we obtain:

\begin{align}
\begin{split}
H_1^{w}(r) &= - \frac{1}{120 \pi \delta} \left\{ \begin{array}{ll}  81 (r/\delta)^7 - 400 (r/\delta)^6 + 735 (r/\delta)^5 - 540 (r/\delta)^4 + 168 (r/\delta)^2  & r \leq \delta, \\ 60 - (\delta/r)^3 - 15 (\delta/r) & r > \delta, \end{array}\right. \quad \textnormal{and}
\\
H_2^{w}(r) &= \frac{1}{120 \pi \delta^3} \left\{ \begin{array}{ll}  63 (r/\delta)^5 - 300 (r/\delta)^4 + 525 (r/\delta)^3 - 360 (r/\delta)^2 + 84  & r \leq \delta, \\ - 3 (\delta/r)^5 + 15 (\delta/r)^3 & r > \delta. \end{array}\right..
\end{split}
\label{eqn:regstokeslet_h1h2_wendland}
\end{align}

For a Gaussian kernel, substituting $G_s$ and $B_s$ from Equation \eqref{eqn:greensfunction_stokesletblob_gaussian} and \eqref{eqn:biharmonicfunction_stokesletblob_gaussian}, we obtain:

\begin{align}
\begin{split}
H_1^{g}(r) &= \frac{1}{8 \pi^{3/2} r^3} \Biggl[ -\sqrt{2} r \sigma \exp \left( -\frac{r^2}{2 \sigma^2} \right) + \sqrt{\pi} (r^2 + \sigma^2) \erf \left( \frac{r}{\sqrt{2} \sigma} \right) \Biggr] \quad \textnormal{and}
\\
H_2^{g}(r) &= \frac{1}{4 \pi r^3 } \Biggl[ \frac{3 \sigma}{\sqrt{2 \pi} r} \exp \left( -\frac{r^2}{2 \sigma^2} \right) + \left( \frac{1}{2} - \frac{3 \sigma^2}{2 r^2} \right) \erf \left( \frac{r}{\sqrt{2} \sigma} \right) \Biggr].
\end{split}
\label{eqn:regstokeslet_h1h2_gaussian}
\end{align}

Here, the limit of these functions as $r \to 0$ are as follows:

\begin{align}
\begin{split}
\lim_{r \to 0} H_1^{g} (r) &= \frac{1}{3 \sqrt{2} \pi^{3/2} \sigma} \quad \textnormal{and}
\\
\lim_{r \to 0} H_2^{g} (r) &= \frac{1}{30 \sqrt{2} \pi^{3/2} \sigma^3}.
\end{split}
\label{eqn:regstokeslet_h1h2_gaussian_limr0}
\end{align}

\subsection{Regularized doublet}

\citet{Cortez2015} provide the doublet kernel as follows:

\begin{equation}
\Delta_{ijk} = \Biggl[ H_2 \delta_{jk} x_i +  \frac{1}{r} \frac{\mathrm{d}H_2}{\mathrm{d}r} x_i x_j x_k + \frac{1}{2} \left( H_2 + \frac{1}{r} \frac{\mathrm{d}H_1}{\mathrm{d}r} \right) (\delta_{ik} x_j + \delta_{ij} x_k) \Biggr] + \frac{1}{2r} \frac{\mathrm{d}G}{\mathrm{d}r} (\delta_{ik} x_j - \delta_{ij} x_k).
\label{eqn:regdoublet_defn}
\end{equation}

Substituting terms for the Wendland kernel, we obtain:

\begin{align}
\begin{split}
\Delta_{ijk}^{w} &= \frac{1}{40 \pi \delta^3} \Biggl[ \frac{x_i x_j x_k}{\delta^2} \left( 105 \left( \frac{r}{\delta} \right)^3 - 400 \left( \frac{r}{\delta} \right)^2 + 525 \left( \frac{r}{\delta} \right) - 240 \right) 
\\
&+  x_k \delta_{ij} \left( -112 + \frac{720 r^2}{\delta^2} - \frac{1225 r^3}{\delta^3} + \frac{800 r^4}{\delta^4} - \frac{189 r^5}{\delta^5} \right) 
\\
&+ (x_j \delta_{ik} + x_i \delta_{jk}) \left( 28 - \frac{120 r^2}{\delta^2} + \frac{175 r^3}{\delta^3} - \frac{100 r^4}{\delta^4} + \frac{21 r^5}{\delta^5} \right)\Biggr] \quad \textnormal{if} \quad r \leq \delta,
\end{split}
\label{eqn:regdoublet_wendland}
\end{align}

\begin{align}
\begin{split}
\Delta_{ijk}^{w} &=  \frac{x_i x_j x_k}{8 \pi r^5} \left( \left( \frac{\delta}{r} \right)^2 - 3 \right) - \frac{x_k \delta_{ij}}{40 \pi r^3} \left( 5 + \left( \frac{\delta}{r} \right)^2 \right) + \frac{x_j \delta_{ik}}{40 \pi r^3} \left( 5 - \left( \frac{\delta}{r} \right)^2 \right)
\\
&- \frac{x_i \delta_{jk}}{40 \pi r^3} \left( \left( \frac{\delta}{r} \right)^2 -5 \right)  \quad \textnormal{if} \quad r > \delta.
\end{split}
\label{eqn:regdoublet_wendland_rgdelta}
\end{align}

Similarly, substituting terms for the Gaussian kernel, we obtain:

\begin{align}
\begin{split}
\Delta_{ijk}^{g} &= \frac{x_i x_j x_k}{4 \pi r^5} \Biggl[ - \frac{(2r^2 + 15 \sigma^2)}{\sqrt{2 \pi} r \sigma} \exp \left( -\frac{r^2}{2 \sigma^2} \right) + \frac{3}{2} \left( \frac{5 \sigma^2}{r^2} - 1 \right) \erf \left( \frac{r}{\sqrt{2} \sigma} \right) \Biggr] 
\\
&+\frac{1}{8 \pi^{3/2} r^5 \sigma} \Biggl[ x_j \delta_{ik} \left( 3 \sqrt{2} r \sigma^2 \exp \left( -\frac{r^2}{2 \sigma^2} \right) + \sqrt{\pi} \sigma (r^2 - 3 \sigma^2) \erf \left( \frac{r}{\sqrt{2} \sigma} \right) \right) 
\\
&+ x_k \delta_{ij} \left(\sqrt{2} r (2r^2 + 3\sigma^2) \exp \left( -\frac{r^2}{2 \sigma^2} \right) - \sqrt{\pi} \sigma (r^2 + 3 \sigma^2) \erf \left( \frac{r}{\sqrt{2} \sigma} \right) \right) \Biggr]
\\
&+\frac{x_i \delta_{jk} }{4\pi r^3}  \Biggl[ \frac{3 \sigma}{\sqrt{2 \pi} r} \exp \left( -\frac{r^2}{2 \sigma^2} \right)  + \left(\frac{1}{2} - \frac{3 \sigma^2}{2 r^2} \right) \erf \left( \frac{r}{\sqrt{2} \sigma} \right) \Biggr].
\end{split}
\label{eqn:regdoublet_gaussian}
\end{align}

\subsection{Regularized dipole}

The dipole velocity field is $u_i = D_{ij} f_j = [D_1 (r) \delta_{ij} + D_2 (r) x_i x_j] f_j$. For a Wendland kernel, $D_1^{w}$ and $D_2^{w}$ are as follows:

\begin{align}
\begin{split}
D_1^{w} &= \frac{1}{20 \pi \delta^3} \left\{ \begin{array}{ll}  147 (r/\delta)^5 - 600 (r/\delta)^4 + 875 (r/\delta)^3 - 480 (r/\delta)^2 + 56  & r \leq \delta, \\ 3 (\delta/r)^5 - 5 (\delta/r)^3 & r > \delta, \end{array}\right. \quad \textnormal{and}
\\
D_2^{w} &= \frac{1}{4 \pi \delta^5} \left\{ \begin{array}{ll}  -21 (r/\delta)^3 + 80 (r/\delta)^2 - 105 (r/\delta) + 48  & r \leq \delta, \\ - (\delta/r)^7 + 3 (\delta/r)^5 & r > \delta. \end{array}\right..
\end{split}
\label{eqn:regdipole_wendland}
\end{align}

For a Gaussian kernel, $D_1^{g}$ and $D_2^{g}$ are as follows:

\begin{align}
\begin{split}
D_1^{g} &= \frac{1}{4 \pi^{3/2} r^5 \sigma} \Biggl[ -\sqrt{2} r (2r^2 + 9 \sigma^2) \exp \left( -\frac{r^2}{2 \sigma^2} \right) + \sqrt{\pi} \sigma (9\sigma^2 - r^2) \erf \left( \frac{r}{\sqrt{2} \sigma} \right) \Biggr] \quad \textnormal{and} \quad
\\
D_2^{g} &= \frac{1}{4 \pi^{3/2} r^7} \Biggl[ \frac{\sqrt{2} r}{\sigma} (2r^2 + 15 \sigma^2) \exp \left( -\frac{r^2}{2 \sigma^2} \right) + 3 \sqrt{\pi} (r^2 - 5 \sigma^2) \erf \left( \frac{r}{\sqrt{2} \sigma} \right) \Biggr].
\end{split}
\label{eqn:regdipole_gaussian}
\end{align}

\subsection{Other elements}

In order to evaluate the Stokeslet image system in Equation \eqref{eqn:imagesystem_stokeslet}, we obtain the difference of rotlets derived from the dipole and Stokeslet blobs, $(\mathcal{R}^{(d)} - \mathcal{R}^{(s)})_{ij}$ as follows:

\begin{equation}
(\mathcal{R}^{(d)} - \mathcal{R}^{(s)})^{w}_{ij} = \frac{\epsilon_{ijk} x_k}{20 \pi \delta^3} \left\{ \begin{array}{ll}  - 84 (r/\delta)^5 + 350 (r/\delta)^4 - 525 (r/\delta)^3 + 300 (r/\delta)^2 - 42  & r \leq \delta, \\ - (\delta/r)^5 & r > \delta. \end{array}\right.
\label{eqn:regrotletdifference_wendland}
\end{equation}

Substituting Equations \eqref{eqn:regstokeslet_h1h2_wendland}, \eqref{eqn:regdoublet_wendland}, and \eqref{eqn:regdipole_wendland}, and \eqref{eqn:regrotletdifference_wendland} into Equation\eqref{eqn:imagesystem_stokeslet} gives us the regularized Stokeslet image system for a Wendland dipole kernel.

Similarly, for a Gaussian kernel, this term is as follows:

\begin{equation}
(\mathcal{R}^{(d)} - \mathcal{R}^{(s)})^{g}_{ij} = \frac{1}{4 \pi^{3/2} r^5 \sigma} \Biggl[ \sqrt{2} r (r^2 + 3 \sigma^2) \exp \left( -\frac{r^2}{2 \sigma^2} \right) - 3 \sqrt{\pi} \sigma^3 \erf \left( \frac{r}{\sqrt{2} \sigma} \right) \Biggr]  \epsilon_{ijk} x_k.
\label{eqn:regrotletdifference_gaussian}
\end{equation}

Substituting Equations \eqref{eqn:regstokeslet_h1h2_gaussian}, \eqref{eqn:regdoublet_gaussian}, \eqref{eqn:regdipole_gaussian}, and \eqref{eqn:regrotletdifference_gaussian} into Equation\eqref{eqn:imagesystem_stokeslet} gives us the regularized Stokeslet image system for a Gaussian dipole kernel.


\section{Taylor-series expansion of the Oseen correction factor}
\label{sec:appendixc_taylorseries_oseencorrection}

We provide a Taylor series expansion of the Oseen correction factor presented in 
Equation \eqref{eqn:oseen_correctionfactor}, as shown below:

\begin{align}
	\begin{split}
	\Psi_{\text{Os}} (\text{Re}_{\delta}) &= 1 - \frac{\text{Re}_{\delta}}{8} + 1.38889 \times 10^{-2} \text{Re}_{\delta}^2 - 1.38889 \times 10^{-3} \text{Re}_{\delta}^3 + 1.26263 \times 10^{-4} \text{Re}_{\delta}^4 
	\\
	&- 1.05219 \times 10^{-5} \text{Re}_{\delta}^5  + 8.09376 \times 10^{-7} \text{Re}_{\delta}^6 - 5.78126 \times 10^{-8} \text{Re}_{\delta}^7 + 3.85417 \times 10^{-9} \text{Re}_{\delta}^8 
	\\
	&- 2.40886 \times 10^{-10} \text{Re}_{\delta}^9 + 1.41697 \times 10^{-11} \text{Re}_{\delta}^{10} - 7.87208 \times 10^{-13} \text{Re}_{\delta}^{11} + 4.1432 \times 10^{-14} \text{Re}_{\delta}^{12} 
	\\
	&- 2.0716 \times 10^{-15} \text{Re}_{\delta}^{13} + 9.86476 \times 10^{-17} \text{Re}_{\delta}^{14} - 4.48398 \times 10^{-18} \text{Re}_{\delta}^{15}.
	\end{split}
	\label{eqn:oseen_correctionfactor_taylorseries}
\end{align}

This expansion is valid for $\text{Re}_{\delta} \leq 10$.

\biboptions{sort&compress}
\bibliographystyle{elsarticle-num-names}


\end{document}